\documentclass[twocolumn]{aastex631}

\usepackage{amsmath}
\usepackage{bm}

\newcommand{\Ra}{R_{\mathrm{a}}}
\newcommand{\Rp}{R_{\mathrm{p}}}

\newcommand{\md}{\mathrm{d}}

\newcommand{\me}{\mathrm{e}}

\newcommand{\bJ}{\bm{J}}

\newcommand{\br}{\bm{r}}
\newcommand{\bk}{\bm{k}}

\newcommand{\btheta}{\bm{\theta}}
\newcommand{\bn}{\bm{n}}

\newcommand{\bOm}{\bm{\Omega}}
\newcommand{\nn}{\nonumber}

\newcommand{\p}{\partial}

\newcommand{\pattern}{\Omega_{\mathrm{p}}}

\newcommand{\barphi}{\widetilde{\phi}}

\newcommand\red[1]{#1}

\begin{document}

\title{Galactokinetics}

\author[0000-0002-5861-5687]{Chris Hamilton}
\affiliation{Institute for Advanced Study, Einstein Drive, Princeton NJ 08540}
\affiliation{Department of Astrophysical Sciences, Princeton University, 4 Ivy Lane, Princeton NJ 08544}

\author[0000-0002-8532-827X]{Shaunak Modak}
\affiliation{Department of Astrophysical Sciences, Princeton University, 4 Ivy Lane, Princeton NJ 08544}

\author[0000-0002-0278-7180]{Scott Tremaine}
\affiliation{Institute for Advanced Study, Einstein Drive, Princeton NJ 08540}

\correspondingauthor{Chris Hamilton} \email{chamilton@ias.edu}

%% Note that the \and command from previous versions of AASTeX is now
%% depreciated in this version as it is no longer necessary. AASTeX 
%% automatically takes care of all commas and "and"s between authors names.

%% AASTeX 6.31 has the new \collaboration and \nocollaboration commands to
%% provide the collaboration status of a group of authors. These commands 
%% can be used either before or after the list of corresponding authors. The
%% argument for \collaboration is the collaboration identifier. Authors are
%% encouraged to surround collaboration identifiers with ()s. The 
%% \nocollaboration command takes no argument and exists to indicate that
%% the nearby authors are not part of surrounding collaborations.

%% Mark off the abstract in the ``abstract'' environment. 
\begin{abstract}
Galactic disks lie at the heart of many astrophysical puzzles. There are sophisticated kinetic theories that  describe some aspects of galaxy disk dynamics, but extracting quantitative predictions from those theories has proven very difficult, meaning they have shed little light on observations/simulations of galaxies. Here, we begin to address this issue by developing a tractable theory describing fluctuations and transport in thin galactic disks. Our main conceptual advance is to split potential fluctuations into asymptotic wavelength regimes relative to orbital guiding radius and epicyclic amplitude (similar to plasma gyrokinetics), and then to treat separately the dynamics in each regime. As an illustration, we apply our results to quasilinear theory, calculating the angular-momentum transport due to a transient spiral. At each stage we verify our formulae with numerical examples. Our approach should simplify many important calculations in galactic disk dynamics. \red{ In a follow-up paper, we apply these ideas to the theory of linear spiral structure in stellar disks.}
\end{abstract}

%% Keywords should appear after the \end{abstract} command. 
%% The AAS Journals now uses Unified Astronomy Thesaurus concepts:
%% https://astrothesaurus.org
%% You will be asked to selected these concepts during the submission process
%% but this old "keyword" functionality is maintained in case authors want
%% to include these concepts in their preprints.
%\keywords{Classical Novae (251) --- Ultraviolet astronomy(1736) --- History of astronomy(1868) --- Interdisciplinary astronomy(804)}

%% From the front matter, we move on to the body of the paper.
%% Sections are demarcated by \section and \subsection, respectively.
%% Observe the use of the LaTeX \label
%% command after the \subsection to give a symbolic KEY to the
%% subsection for cross-referencing in a \ref command.
%% You can use LaTeX's \ref and \label commands to keep track of
%% cross-references to sections, equations, tables, and figures.
%% That way, if you change the order of any elements, LaTeX will
%% automatically renumber them.
%%
%% We recommend that authors also use the natbib \citep
%% and \citet commands to identify citations.  The citations are
%% tied to the reference list via symbolic KEYs. The KEY corresponds
%% to the KEY in the \bibitem in the reference list below. 

%%%%%%%%%%%%%%%%%%%%%%%%%%%%%%%%%%%%%%%%
%%%%%%%%%%%%%%%%%%%%%%%%%%%%%%%%%%%%%%%%
\section{Introduction}
\label{sec:Introduction}
%%%%%%%%%%%%%%%%%%%%%%%%%%%%%%%%%%%%%%%%
%%%%%%%%%%%%%%%%%%%%%%%%%%%%%%%%%%%%%%%%

Galactic disks occupy an especially prominent place in modern astrophysics, as the setting for many of our most profound puzzles. The nature of dark matter is still a mystery half a century after the discovery of flat rotation curves \citep{mo2010galaxy}. Our understanding of galaxy formation in the early Universe is being upended by the observation of very massive early disk galaxies by JWST \citep{wu2022identification, Wang2025-disk}. The assembly history and evolution of the Milky Way is still unknown although its fingerprints are surely embedded in the exquisite GAIA data \citep{deason2024galactic,hunt2025milky}. For these and other problems to be solved, we need an improved understanding of the dynamics and evolution of galactic disks.

The theorist's picture of a galactic disk is a collection of stars on nearly circular, nearly coplanar orbits, which exchange energy and angular momentum with a bath of potential `fluctuations'. These fluctuations may comprise spiral transients, a central bar, dark matter substructure, interstellar medium (ISM) gas structures, and so on. Given the importance of galaxy disk evolution, it is highly desirable to develop kinetic theories that describe this energy and angular-momentum exchange accurately. Indeed, a great deal of sophisticated formalism has been developed for just this purpose, beginning several decades ago \citep{julian1966non, kalnajs1971dynamics, Lynden-Bell1972-ve}, long before observations of galaxy disks were of high enough quality to demand it, and long before the evolution of such disks could be simulated accurately on computers. For an overview, see \cite{hamilton2024kinetic}.

\red{Yet it proves difficult to connect} abstract kinetic theory to either observations or numerical simulations. Milky Way astronomy is a prime example. \red{Surveys like} GAIA \red{and APOGEE have} provided \red{an intricate} kinematic and chemical map of the Galaxy within a few kpc of the Sun, and several puzzles have thereby arisen regarding the formation of the vertical phase space `snail' \citep{antoja2018dynamically}, the remarkably efficient radial migration of disk stars \citep{Frankel2020-vy}, and so on. Each of these puzzles has inspired various solutions, which normally consist of a plausible set of initial conditions and dynamical perturbations that are forward-modeled numerically, but in very few cases has this approach yielded a convincing answer that excludes all others: the parameter space of possibilities is simply too large for such models to land a knockout blow. But analytic (or semi-analytic) kinetic theory has not really helped either. With a few exceptions (e.g., \citealt{Chiba2021-cc,banik2022comprehensive,widrow2023swing}), kinetic theorists have been unable to write down tractable expressions --- or even robust estimates or scaling laws --- that can be meaningfully compared against observation.   In short, the quality of modern observations demands a tractable, \textit{predictive} kinetic theory of fluctuations and transport in galactic disks, which is currently lacking.

The issue is that, although the formal kinetic equations describing disk evolution are (we think) largely known, extracting quantitative predictions from those equations is very difficult. 
For instance, one must almost always evaluate the $(2d+1)$-dimensional function
\begin{equation}
    \delta \phi_{\bn}(\bJ, t),
    \label{eqn:deltaphinJt}
\end{equation}
where $d$ is the number of spatial dimensions ($d=2$ for a razor-thin disk, which is what we will concentrate on here).
Really this is just the potential fluctuation $\delta \phi(\br,t)$, except that it is (i) first transformed from position \red{space} ($\br$) to angle-action space ($\btheta, \bJ$), and (ii) then (Fourier) transformed from angle-space to wavenumber ($\bn$) space. Furthermore, in many applications this function then undergoes (iii) a projection onto a set of biorthogonal basis functions, and then possibly (iv) yet another (Fourier or Laplace) transform from the time domain into the frequency domain, along with any necessary analytic continuation. After further manipulations there may even follow (v) an inverse Fourier or Laplace transform back to the time domain, likely requiring contour integration. This done, one typically must (vi) sum over all possible wavenumbers $\bn$, of which there are in principle infinitely many. These devilish layers of additional complexity mean that in practice \red{it can be} computationally cheaper to simulate a system's entire history directly than to evaluate an approximate `analytic' expression designed to foretell its behavior over a single timestep\footnote{\red{This issue is well illustrated by \cite{roule2025long}, who (successfully, but painstakingly) predicted the very early evolution of a statistical ensemble of 2D Mestel disks using Balescu-Lenard kinetic theory.}}. \red{This tends to limit the usefulness of the kinetic approach.}

\red{Broadly speaking, there are two ways to address this issue. One is to develop ever more efficient numerical algorithms that can accurately evaluate \eqref{eqn:deltaphinJt} and similar functions. This is the idea (or one of the ideas) behind the `torus mapping' method pioneered by \cite{binney2016torus,Binney2018orbital,binney2020angle} and the linear response package developed by \cite{petersen2024predicting}. In this series of papers we will pursue an alternative --- and complementary --- analytic approach, which
necessarily sacrifices some of the accuracy of the aforementioned numerical methods
but offers greater analytic insight and interpretability.}

The key \red{idea} is that what we really care about is not the potential fluctuations themselves, but rather how they couple to orbits. In the case of galactic disks, there are two key orbital length scales --- the guiding center radius $R_\mathrm{g}$ and the typical epicyclic amplitude $a$,  with $a\ll R_\mathrm{g}$ in the common case of a `cool' disk --- and the potential fluctuations $\delta \phi(\br, t)$ can be classified relative to these length scales. As we will show, in the resulting asymptotic regimes, one can derive expressions for $\delta \phi_{\bn}(\bJ, t)$ that are drastically simpler than in the general case referred to above, but also remarkably accurate in practice. Moreover, for some calculations the fluctuations from one asymptotic regime completely dominate the orbital couplings, so complicated kinetic equations can be greatly simplified. Our approach is reminiscent of the gyrokinetic theory of plasmas, in which the `mean-field' orbits of particles consist of Larmor motion around magnetic field lines and force fluctuations are categorized relative to the Larmor radius. By analogy, we give our framework the name \textit{galactokinetics}.

The purpose of this paper is to develop the galactokinetic approximation scheme and show how it renders tractable a toy problem in the (quasilinear) kinetic theory of galactic disk dynamics. While some --- perhaps even most --- of the equations we develop below can be found scattered across the galaxy dynamics literature \citep{shu1970density,Lynden-Bell1972-ve,Mark1974-jj,Carlberg1985-gf,Sygnet1988-eq}, the relevant approximation scheme does not seem to have been developed systematically before, nor has it been tested numerically. \red{In \cite{GK2} --- hereafter Paper II --- we apply the formal techniques we develop here to construct a unified linear theory of spiral structure in stellar disks.} %In Paper III we will apply them to the problem of radial heating and migration in galactic disks.}
%problem of the heating of the Milky Way disk.

Our plan for the rest of this work is as follows. First we introduce our basic model of a thin galactic disk (\S\ref{sec:model}) and discuss the epicyclic approximation (\S\ref{sec:Approximations}). We then classify potential fluctuations according to their wavelengths, derive explicit expressions for $\delta \phi_{\bn}(\bJ, t)$ in each class, and validate the expressions numerically (\S\ref{sec:Potential_Section}). Having done this, we apply our asymptotic results to quasilinear kinetic theory (\S\ref{sec:kinetic_theory}). We conclude in \S\ref{sec:Discussion}.

\medskip

%%%%%%%%%%%%%%%%%%%%%%%%%%%%%%%%%%%%%%%%
%%%%%%%%%%%%%%%%%%%%%%%%%%%%%%%%%%%%%%%%
\section{Model}
\label{sec:model}
%%%%%%%%%%%%%%%%%%%%%%%%%%%%%%%%%%%%%%%%
%%%%%%%%%%%%%%%%%%%%%%%%%%%%%%%%%%%%%%%%

We consider a razor-thin stellar disk made up of equal-mass stars (and possibly embedded in a rigid dark matter halo). We will describe the disk using polar coordinates $\br = (\varphi, R)$, and hence velocities using $\bm{v} = (v_\varphi, v_R) = (R\dot{\varphi}, \dot{R})$. For simplicity, we will ignore any `vertical' motion (motion normal to the disk plane).

To begin our construction, at a given time let the \textit{exact} gravitational potential \red{at location $\br$} (including external perturbations, if they are present) be $\phi(\br) = \phi(\varphi, R)$. Then the exact Hamiltonian of a test star is 
\begin{equation}
    \label{eq:Exact_Ham}
    h(\br, \bm{v}) = \frac{1}{2} \bm{v}^2 + \phi(\br).
\end{equation}
The azimuthal average of $h$ is 
\begin{equation}
    \label{eq:Axi_Ham}
    \widetilde{h}(R, \bm{v}) \equiv \frac{1}{2} \bm{v}^{2} + \barphi(R),
\end{equation}
where for any function $q(\br)$ we define
\begin{equation}
    \widetilde{q}(R) \equiv \int_0^{2\pi} \frac{\md \varphi}{2\pi} \, q(\br).
\end{equation}
We have not yet stipulated where the Hamiltonian $h$ comes from; it might be generated self-consistently by the stars, or externally imposed, or some combination of these.

We suppose that $\phi$ is reasonably close to axisymmetry, i.e., not far from $\barphi$, and we can construct angle-action variables $(\btheta, \bJ)$ such that the associated Hamiltonian $\widetilde{h}$ only depends on $\bJ$. These actions are $\bJ = (J_\varphi, J_R)$ where 
\begin{equation}
    J_\varphi = Rv_\varphi,
\end{equation}
is the angular momentum, and
\begin{equation}
    \label{eq:def_Jr}
    J_{R} = \frac{1}{\pi} \int_{\Rp}^{\Ra} \md R \, v_{R}(R),
\end{equation}
is the radial action. Here,
\begin{equation}
v_R(R) \equiv \sqrt{2[E - \barphi(R)] - J_\varphi^2/R^2},
\end{equation}
is the radial velocity, and $R_\mathrm{p/a}$ are the peri/apocenter radii, i.e., the solutions to $v_R(R)=0$. The conjugate angles $\btheta = (\theta_\varphi, \theta_R)$ are then the azimuthal angle
\begin{equation}
    \label{eq:azimuthal_angle}
    \theta_\varphi = \varphi + \int_C \md R' \, \frac{\Omega_\varphi - J_\varphi/R'^2 }{v_R(R')},
\end{equation}
and the radial angle 
\begin{equation}
    \label{eq:radial_angle}
    \theta_R = \Omega_R \int_C \frac{\md R'}{v_R(R')},
\end{equation}
where
\begin{equation}
    \bOm(\bJ) = (\Omega_\varphi, \Omega_R) \equiv \frac{\partial \widetilde{h}}{\partial \bJ},
    \label{eqn:frequency_definition}
\end{equation}
are the orbital frequencies, and $C$ is the contour going from the pericenter $\Rp$ up to the current position $R(\theta_R, \bJ)$, along the radial oscillation. Thus $\theta_R$ increases by $2\pi$ in one radial period $ 2\pi/\Omega_R,$ while \red{in the same time} $\theta_\varphi$ increases by $2\pi\Omega_\varphi/\Omega_R$.

%%%%%%%%%%%%%%%%%%%%%%%%%%%%%%%%%%%%%%%%
%%%%%%%%%%%%%%%%%%%%%%%%%%%%%%%%%%%%%%%%
\subsection{Mean field and fluctuations}
\label{sec:mean_fluctuations}
%%%%%%%%%%%%%%%%%%%%%%%%%%%%%%%%%%%%%%%%
%%%%%%%%%%%%%%%%%%%%%%%%%%%%%%%%%%%%%%%%

Having constructed our angle-action variables, we can express any function on phase space in terms of them via $\br = \br(\btheta, \bJ)$ and $\bm{v} = \bm{v}(\btheta, \bJ)$. In particular, we can return to our original exact Hamiltonian $h$ and split it into a $\btheta$-independent `mean-field' part and a $\btheta$-dependent `fluctuation':
\begin{equation}
    \label{eq:h1}
    h(\br(\btheta, \bJ), \bm{v}(\btheta, \bJ))  = h_0(\bJ) + \delta h(\btheta, \bJ),
\end{equation}
where the $\btheta$-average of some quantity $q(\btheta, \bJ)$ is
\begin{equation}
    \label{eq:Definiton_Angle_Average}
    q_0(\bJ) \equiv \int \frac{\md \btheta}{(2\pi)^2} \, q(\btheta, \bJ).
\end{equation}
Here, the `$0$' subscript reminds us that $q_0$ is equivalent to the zeroth Fourier mode of $q$ when expanded in angles (see, e.g., equation \eqref{eq:Fourier_deltaf} below). Note also that here and henceforth we are suppressing the time argument of the fluctuations. It is easy to show that the `mean-field Hamiltonian' $h_0$ and the `axisymmetric Hamiltonian' $\widetilde{h}$ are equivalent, and that 
\begin{equation}
    \label{eq:Non_Axi_phi}
    \delta h(\btheta, \bJ) = \phi(\br) - \barphi(R) \equiv \delta \phi(\btheta, \bJ),
\end{equation}
i.e., the fluctuation part of the Hamiltonian is just the non-axisymmetric part of the potential.\footnote{\red{These properties are consequences of the fact that we \textit{chose} to construct our angle-action variables using $\widetilde{\phi}$.
We could just as well have constructed them with reference to some different axisymmetric potential,
but then the $\varphi$-averaged and $\btheta$-averaged Hamiltonians $\widetilde{h}(R,\bm{v})$, $h_0(\bJ)$ would generally not be equivalent, and the angle-dependent Hamiltonian $\delta h(\btheta, \bJ)$ would contain an axisymmetric piece (depending on angle through $\theta_R$).
Indeed, this is what one does in practice if, e.g., one considers epicyclic orbits in a smooth disk and then adds an axisymmetric ring or groove to the potential. This subtlety will be of no importance for the rest of this paper so we cease to mention it hereafter.}} Hence, from now on we write
\begin{equation}
    \label{eq:h}
    h(\br(\btheta, \bJ), \bm{v}(\btheta, \bJ))  = h_0(\bJ) + \delta \phi(\btheta, \bJ),
\end{equation}
and refer to $h_0$ and $\delta \phi$ as the mean-field Hamiltonian and the potential perturbation respectively.

Another key quantity we must introduce is the \textit{guiding center potential perturbation}, which is defined as 
\begin{equation}
    \label{eq:Potential_Fluctuation_Guiding_Center}
    \delta \Phi(\theta_\varphi, J_\varphi) \equiv \delta \phi(\theta_\varphi, \theta_R, J_\varphi, 0).
\end{equation}
i.e., the potential fluctuation {evaluated at the instantaneous location of the star's guiding center}. Note that $\delta \Phi$ is necessarily independent of $\theta_R$. 
%Of course, this is only equivalent to $\delta \phi$ for perfectly circular orbits.

%%%%%%%%%%%%%%%%%%%%%%%%%%%%%%%%%%%%%%%%
%%%%%%%%%%%%%%%%%%%%%%%%%%%%%%%%%%%%%%%%
\subsection{Distribution function}
\label{sec:distribution_function}
%%%%%%%%%%%%%%%%%%%%%%%%%%%%%%%%%%%%%%%%
%%%%%%%%%%%%%%%%%%%%%%%%%%%%%%%%%%%%%%%%

Let the exact distribution function (DF) of stars be $f(\btheta, \bJ, t)$, normalized such that 
\begin{equation}
    \int \md\btheta\, \md\bJ \, f(\btheta, \bJ, t) = M,
\end{equation}
the total mass in stars. This DF satisfies the equation
\begin{equation}
    \label{eq:Klimontovich}
    \frac{\partial f}{\partial t} + \frac{\partial f}{\partial \btheta}\cdot   \frac{\partial h}{\partial \bJ} - \frac{\partial f}{\partial \bJ}\cdot   \frac{\partial h}{\partial \btheta} = 0.
\end{equation}
Without approximation we can always expand $f$ like we expanded $h$ in equation \eqref{eq:h}:
\begin{equation}
    f(\btheta, \bJ) = f_0(\bJ) + \delta f(\btheta, \bJ).
    \label{eqn:f_f0_deltaf}
\end{equation}
(Recall that a subscript zero refers to an angle-average, viz.\ equation \eqref{eq:Definiton_Angle_Average}.)
Plugging these expansions into \eqref{eq:Klimontovich} gives
\begin{align}
    \label{eq:Fluctuation_Evolution}
    &\frac{\partial \delta f}{\partial t} + \bOm \cdot \frac{\partial \delta f}{\partial \btheta} - \frac{\partial f_0}{\partial \bJ} \cdot \frac{\partial \delta \phi}{\partial \btheta}
    \nn
    \\
    &= -\frac{\partial}{\partial \btheta} \cdot \left( \delta f \frac{\partial \delta \phi}{\partial \bJ} \right) + \frac{\partial}{\partial \bJ} \cdot \left[ \delta f \frac{\partial \delta \phi}{\partial \btheta} - \left( \delta f \frac{\partial \delta \phi}{\partial \btheta}\right)_0 \, \right],
\end{align}
and
\begin{equation}
    \label{eq:Axisymmetric_Evolution}
    \frac{\partial f_0}{\partial t}  = \frac{\partial}{\partial \bJ} \cdot \left(\delta f \frac{\partial \delta \phi}{\partial \btheta}\right)_0.
\end{equation}

We will frequently be interested in \textit{marginalized} functions which do not depend on radial angle or radial action.  To this end we introduce an overline notation to denote an integral (\textit{not} an average) over $\theta_R$ and $J_R$:
\begin{equation}
    \overline{q} \equiv  \int_0^\infty \md J_R \int_0^{2\pi} \md \theta_R \, q(\btheta, \bJ).
\end{equation}
The most important function we wish to marginalize is the DF $f$. Integrating both sides of \eqref{eqn:f_f0_deltaf} with respect to $\theta_R$ and $J_R$, we have
\begin{equation}
    F(\theta_\varphi, J_\varphi) = F_0(J_\varphi) + \delta F(\theta_\varphi, J_\varphi),
\end{equation}
where
\begin{align}
    F(\theta_\varphi, J_\varphi) & \equiv \overline{ f(\btheta, \bJ)}, \\
        \label{eq:Marginalized_Mean_DF}
    F_0(J_\varphi) & \equiv \overline{ f_0(\bJ) } = 2\pi \int_0^\infty \md J_R \, f_0(\bJ), \\
    \label{eq:Marginalized_Fluctuating_DF}
    \delta F(\theta_\varphi, J_\varphi) & \equiv \overline{ \delta f(\btheta, \bJ) }.
\end{align}
By applying the same procedure to equations \eqref{eq:Fluctuation_Evolution}--\eqref{eq:Axisymmetric_Evolution}, we find that $\delta F$ and $F_0$ evolve according to 
\begin{align}
    \frac{\partial \delta F}{\partial t}  + \overline{ \left( \Omega_\varphi  \frac{\partial \delta f}{\partial \theta_\varphi} \right)} &- \overline{ \left( \frac{\partial \delta \phi}{\partial \theta_\varphi} \frac{\partial f_0}{\partial J_\varphi}\right)} \nn
    \\
    &=  - \frac{\p}{\p \theta_\varphi} \delta \mathcal{G} - \frac{\partial}{\partial J_\varphi} \delta \mathcal{F},
     \label{eq:Fluctuation_Evolution_1D}
\end{align}
and
\begin{equation}
    \label{eq:Mean_Evolution_1D}
    \frac{\partial F_0}{\partial t}  = -\frac{\partial}{\partial J_\varphi} \mathcal{F}_0,
\end{equation}
respectively, where
\begin{equation}
    \mathcal{F}(\theta_\varphi, J_\varphi) \equiv - \overline{ \left(\delta f \frac{\partial \delta \phi}{\partial \theta_\varphi}\right)},
\end{equation}
\begin{equation}
     \mathcal{G}(\theta_\varphi, J_\varphi) \equiv \overline{ \left(\delta f \frac{\partial \delta \phi}{\partial J_\varphi} \right)},    
     \label{eqn:Curly_G}
\end{equation}
and $\delta \mathcal{G} = \mathcal{G} - \mathcal{G}_0$, $\delta \mathcal{F} = \mathcal{F} - \mathcal{F}_0$ are the $\theta_\varphi$-dependent parts of $\mathcal{G}$ and $\mathcal{F}$. We emphasize that equations \eqref{eq:Fluctuation_Evolution_1D}--\eqref{eqn:Curly_G} are exact; we have made no approximations so far.
%\SM{Possibly move this last sentence to the beginning of section 3.1?}

%%%%%%%%%%%%%%%%%%%%%%%%%%%%%%%%%%%%%%%%
%%%%%%%%%%%%%%%%%%%%%%%%%%%%%%%%%%%%%%%%
\subsection{Fourier expansion}
%%%%%%%%%%%%%%%%%%%%%%%%%%%%%%%%%%%%%%%%
%%%%%%%%%%%%%%%%%%%%%%%%%%%%%%%%%%%%%%%%

For explicit calculations it will be useful for us to introduce Fourier expansions of $\delta f$ and $\delta \phi$:
\begin{align}
    \label{eq:Fourier_deltaf}
    \delta f(\btheta, \bm{J}) = & \sum_{{\bm{n}}} \delta f_{{\bm{n}}}(\bm{J}) \, \me^{i {\bm{n}}\cdot\btheta}, \\
    \label{eq:Fourier_deltaphi}
    \delta \phi(\btheta, \bm{J}) = & \sum_{{\bm{n}}} \delta \phi_{{\bm{n}}}(\bm{J}) \, \me^{i {\bm{n}}\cdot\btheta},
\end{align}
where $\bn=(n_\varphi, n_R)$ are vectors of integers. In particular, various terms in equations \eqref{eq:Fluctuation_Evolution_1D}--\eqref{eqn:Curly_G} can then be expressed as Fourier series; some explicit expressions for these are given in Appendix \ref{sec:Explicit_Fourier}.
The most important one is
\begin{align}
    \mathcal{F}_0(J_\varphi) = 2\pi \sum_{\bn} i n_\varphi \int_0^\infty \md J_R \, \delta f_{\bn}(\bJ) \delta\phi^*_{\bn}(\bJ).
    \label{eqn:phi_flux_Fourier}
\end{align}
We will also benefit from developing the functions $\delta F$ and $\delta \Phi$ as Fourier series in the angle $\theta_\varphi$:
\begin{align}
    \label{eq:fourier_marginalized_DF}
    \delta F(\theta_\varphi, J_\varphi) & = \sum_{n_\varphi} \me^{in_\varphi \theta_\varphi}\delta F_{n_\varphi}(J_\varphi), \\
    \label{eq:fourier_guidingcenter_phi}
    \delta \Phi(\theta_\varphi, J_\varphi) & = \sum_{n_\varphi} \me^{in_\varphi \theta_\varphi}\delta \Phi_{n_\varphi}(J_\varphi).
\end{align}
The coefficients in these expansions are related 
to those in \eqref{eq:Fourier_deltaf}--\eqref{eq:Fourier_deltaphi} by 
\begin{align}
    \label{eq:DF_Fluctuation_Fourier_Identify}
   & \delta F_{n_\varphi}(J_\varphi) = 2\pi \int_0^\infty \md J_R \, \delta f_{(n_\varphi, 0)}(J_\varphi, J_R),
    \\
    &\delta \Phi_{n_\varphi}(J_\varphi) = \delta\phi_{(n_\varphi, 0)}(J_\varphi, 0).
    \label{eq:Potential_Fluctuation_Fourier_Identify}
\end{align}

\medskip

%%%%%%%%%%%%%%%%%%%%%%%%%%%%%%%%%%%%%%%%
%%%%%%%%%%%%%%%%%%%%%%%%%%%%%%%%%%%%%%%%
\section{Epicyclic approximation}
\label{sec:Approximations}
%%%%%%%%%%%%%%%%%%%%%%%%%%%%%%%%%%%%%%%%
%%%%%%%%%%%%%%%%%%%%%%%%%%%%%%%%%%%%%%%%

The results of \S\ref{sec:model} are exact, but to render the theory tractable we need to make some approximations. The key approximation is that the disk is cool (\S\ref{sec:Cool_DF}) which means that mean-field orbits are well-described by epicycles (\S\ref{sec:epicyclic}).

%%%%%%%%%%%%%%%%%%%%%%%%%%%%%%%%%%%%%%%%
%%%%%%%%%%%%%%%%%%%%%%%%%%%%%%%%%%%%%%%%
\subsection{Cool distribution function}
\label{sec:Cool_DF}
%%%%%%%%%%%%%%%%%%%%%%%%%%%%%%%%%%%%%%%%
%%%%%%%%%%%%%%%%%%%%%%%%%%%%%%%%%%%%%%%%

We will assume throughout this study that our galactic disk is \textit{cool}, by which we mean that at any given time,  the underlying mean-field DF $f_0(\bJ)$ is heavily biased towards near-circular orbits, i.e., those with $J_R \ll J_\varphi$. More precisely, we define the expectation value $\langle q \rangle$ of any quantity $q(\btheta, \bJ)$ as
  \begin{equation}
    \langle q \rangle = \frac{1}{M} \int \md \btheta \, \md \bJ \, f(\btheta, 
    \bJ) \, q(\btheta, \bJ).
    \label{eqn:brackets}
\end{equation}
\red{In fact, we often consider such expectation values localized \red{near} some value of $J_\varphi$; then $f$ and $M$ should be interpreted as the DF and \red{total} stellar mass only of those stars with angular momenta in the vicinity of $J_\varphi$.}

The statement that the disk is cold is then equivalent to the statement that, at all relevant angular momenta $J_\varphi$, 
\begin{equation}
    \epsilon(J_\varphi) \equiv \left( \frac{\langle J_R\rangle}{J_\varphi}\right)^{1/2} \ll 1.
    \label{eqn:def_epsilon}
\end{equation}
Usually we will drop the $J_\varphi$ dependence of the small quantity \eqref{eqn:def_epsilon} and just refer to it as $\epsilon$.

How small is $\epsilon$ in reality? Recently, \cite{binney2023self} fit action-based DFs to various components of our Galaxy using GAIA data. They separated the Galactic disk into young, medium, old and thick components, each of which had a DF whose radial action dependence took the form $f_0 \propto \exp(-J_R/\langle J_R\rangle)$, where $\langle J_R \rangle$ depends on $J_\varphi$. Using their best-fit models we deduce that at the Solar radius (assuming $J_\varphi = 8\,\mathrm{kpc} \times 220\, \mathrm{km\,s}^{-1}$) these component disks have 
\begin{equation}
    \epsilon(8\, \mathrm{kpc} \times 220 \, \mathrm{km\,s}^{-1}) =  [0.11, 0.13, 0.13, 0.17].
\end{equation}
Alternatively, evaluating $\epsilon$ somewhat closer to the Galactic center, we have 
\begin{equation}
    \epsilon(4\, \mathrm{kpc} \times 220 \, \mathrm{km\,s}^{-1}) =  [0.12, 0.16, 0.18, 0.25],
\end{equation}
or somewhat further from the Galactic center, we have
\begin{equation}
    \epsilon(12\, \mathrm{kpc} \times 220 \, \mathrm{km\,s}^{-1}) =  [0.10, 0.11, 0.11, 0.14].
\end{equation}
Thus, at least for \red{most} regions of interest in the Milky Way, higher order ($\mathcal{O}(\epsilon^2)$) \red{relative} corrections to the theory we are about to develop are at \red{the level of a few percent at most}.

%%%%%%%%%%%%%%%%%%%%%%%%%%%%%%%%%%%%%%%%
%%%%%%%%%%%%%%%%%%%%%%%%%%%%%%%%%%%%%%%%
\subsection{Epicyclic orbits}
\label{sec:epicyclic}
%%%%%%%%%%%%%%%%%%%%%%%%%%%%%%%%%%%%%%%%
%%%%%%%%%%%%%%%%%%%%%%%%%%%%%%%%%%%%%%%%

When $\epsilon \ll 1$, most orbits are well described using the epicyclic approximation. The \red{canonical mapping to angle-action coordinates for} epicyclic orbits in a central potential $\barphi(R)$ is \red{\citep{Binney2008-ou}}:
\begin{align}
    \label{eq:Epicycle_Azimuthal_Angle}
    \varphi &= \theta_\varphi + \gamma
     \frac{a_R}{R_\mathrm{g}}\sin\theta_R \red{+ \frac{1}{4}a_R^2\frac{\md \kappa}{\md J_\varphi}
     \sin 2\theta_R,} \\
    \label{eq:Epicycle_Radius}
    R &= R_\mathrm{g} -  {a_R} \cos \theta_R, \\
    \label{eq:epicycle_Jphi}
    J_\varphi &= R_\mathrm{g}^2 \, \red{\Omega}, \\
    \label{eq:epicycle_Jr}
    J_R &= \frac{E_R}{\red{\kappa}}.
\end{align}
\red{Here,} $R_\mathrm{g}(J_\varphi)$ is the guiding radius (the radius of a circular orbit of angular momentum $J_\varphi$), $a_R$ is the amplitude of the radial epicycle:
\begin{equation}
    \label{eq:amplitudes_epicycle}
    a_R(\bJ) \equiv \frac{\sqrt{2E_R}}{\red{\kappa}},
\end{equation}
$E_R$ is the radial energy:
\begin{equation}
   E_R(\bJ) \equiv \frac{1}{2}v_R^2 + \frac{1}{2}\red{\kappa}^2 (R-R_\mathrm{g})^2,
\end{equation}
\red{$\Omega$, $\kappa$ are the azimuthal and radial frequencies of a circular orbit with radius $R_\mathrm{g}$:
\begin{align}
    \label{eq:Circular_Frequency}
    \Omega^2(J_\varphi) &\equiv \left(\frac{1}{R} \frac{\partial \barphi}{\partial R}\right)_{R = R_\mathrm{g}} \\
    \label{eq:Radial_Epicyclic_Frequency}
    {\kappa}^2(J_\varphi) &\equiv \left(\frac{2\Omega}{R} \frac{\partial(\Omega R^2)}{ \partial R}\right)_{R = R_\mathrm{g}},
\end{align}}
and $\gamma$ is an order-unity dimensionless number:
\begin{equation}
    \gamma(J_\varphi) \equiv \frac{2\Omega}{\kappa}.
\end{equation}
The azimuthal period is then approximately $T_\varphi = 2\pi/\Omega$, and the radial period is $T_R=2\pi/\kappa$. (For reference, in a flat rotation curve, we have $\kappa = \sqrt{2}\Omega$ and hence $\gamma = \sqrt{2}$.) 

\red{Note that this} epicyclic mapping is accurate up to corrections $\mathcal{O}(\epsilon_R^2)$,
where
\begin{align}
    \label{eq:epsilon_R}
    \epsilon_R(\bJ) &\equiv \left( \frac{J_R}{J_\varphi} \right)^{1/2} =\frac{1}{\sqrt{\gamma}} \frac{a_R}{R_\mathrm{g}},
\end{align}
is a $\bJ$-dependent small quantity, whose \red{rms value is $\sqrt{\langle \epsilon_R^2 \rangle} = \epsilon$} (see equation \eqref{eqn:def_epsilon}).
\red{However, unlike most treatments of the epicyclic approximation, 
we have included a third term in \eqref{eq:Epicycle_Azimuthal_Angle} which is also $\mathcal{O}(\epsilon_R^2)$. 
This term reflects the fact that the frequency of azimuthal motions depends on orbital eccentricity \citep{Grivnev1988}; at the 
population level this is the cause of asymmetric drift \citep{Binney2008-ou}. Moreover, it was pointed out by \cite{dehnen1999approximating} that this term must be included in order to make the mapping from $(\br,\bm{v})\to(\btheta,\bJ)$ canonical\footnote{\red{Note, though, that the factor $1/4$ is missing in \cite{dehnen1999approximating}'s equation (20f) (c.f. equation (3.265) of \citealt{Binney2008-ou}).}}.
More precisely, the canonical mapping generates the azimuthal and radial frequencies (equation \eqref{eqn:frequency_definition}):
\begin{align}
    \Omega_\varphi &= \Omega  + \Omega_\mathrm{D},
    \label{eqn:azi_freq}
    \\
    \Omega_R &= \kappa,
    \label{eqn:rad_freq}
\end{align}
the first of which differs from the circular frequency $\Omega$ due to
an additional drift term
   \begin{equation}
    \Omega_\mathrm{D}(\bJ) \equiv \frac{\md \kappa}{\md J_\varphi} J_R,
    \label{eqn:Dehnen}
\end{equation}
\red{which is typically $\sim \epsilon_R^2 \Omega$.}
We will refer to it as the \textit{Dehnen drift} throughout this paper.}

It is worth noting the relation between $\langle J_R \rangle$ and the mean epicyclic amplitude and radial velocity dispersion.
Supposing $f$ is restricted to a narrow enough range of $J_\varphi$ values such that $\kappa$ is roughly constant, the root-mean-square epicyclic amplitude is
\begin{equation}
    a \equiv \sqrt{\langle a_R^2\rangle } = \left( \frac{2 \langle J_R\rangle}{\kappa}\right)^{1/2}.
    \label{eqn:rms_epicyclic_amplitude}
\end{equation}
Meanwhile, if $f$ is completely phase-mixed (i.e., it has no $\btheta$-dependence) then we can also straightforwardly relate $\langle J_R \rangle$ to the root-mean-square radial velocity:
\begin{equation}
    \sigma \equiv \sqrt{\langle v_R^2\rangle } =\left(\kappa \langle J_R\rangle \right)^{1/2}.
    \label{eqn:rms_radial_velocity}
\end{equation}
We can also relate $a$ and $\sigma$ to the small parameter \eqref{eqn:def_epsilon}:
\begin{align}
    \epsilon &=  \sqrt{\langle \epsilon_R^2\rangle } =  \frac{1}{\sqrt{\gamma}}\frac{a}{R_\mathrm{g}} =  \sqrt{\frac{\gamma}{2}}\frac{\sigma}{V} 
    \\
    &\simeq 0.076 \times  \left(\frac{\gamma}{\sqrt{2}}\right)^{1/2}\left( \frac{\sigma}{20\, \mathrm{km\,s}^{-1}} \right)\left( \frac{V}{220\, \mathrm{km\,s}^{-1}} \right)^{-1},
\end{align}
where $V = R_\mathrm{g}\Omega$ is the circular speed.

\begin{table}
\centering
\caption{Properties of the two orbits shown in Figure \ref{fig:three_orbits_rotating}.  Quantities below the single horizontal line are computed using the epicyclic approximation (\S\ref{sec:epicyclic}).\label{table:three_orbits}}
\begin{tabular}{ccccc}
 Orbit & &(a) &(b)\\
\hline
\hline
$\Rp/R_0$ & & $7.5$ & $7.0$  \\
$\Ra/R_0$ & & $8.5$ & $9.0$  \\
$J_\varphi/(R_0V_0)$ & & $7.97$ & $7.90$   \\
$\mathrm{sma}$ & & $8.0$ & $8.0$  \\
$\mathrm{ecc}$ & & $0.063$ & $0.13$   \\
%$R_\mathrm{g}/R_0$ & $7.97$ & $7.90$   \\
\hline
$T_\varphi/T_0$& & $50.1$ & $49.6$   \\
$\Omega T_0$ & & $0.13$ & $0.13$  \\
$J_R/(R_0V_0) $ & & $0.022$ & $0.088$ \\
$a_R/R_0$ & & $0.50$ & $0.99$ \\
$\epsilon_R$ & & $0.053$ & $0.11$  \\
\hline
\end{tabular}
\end{table}
\begin{figure}
    \centering
    \includegraphics[width=0.49\textwidth]{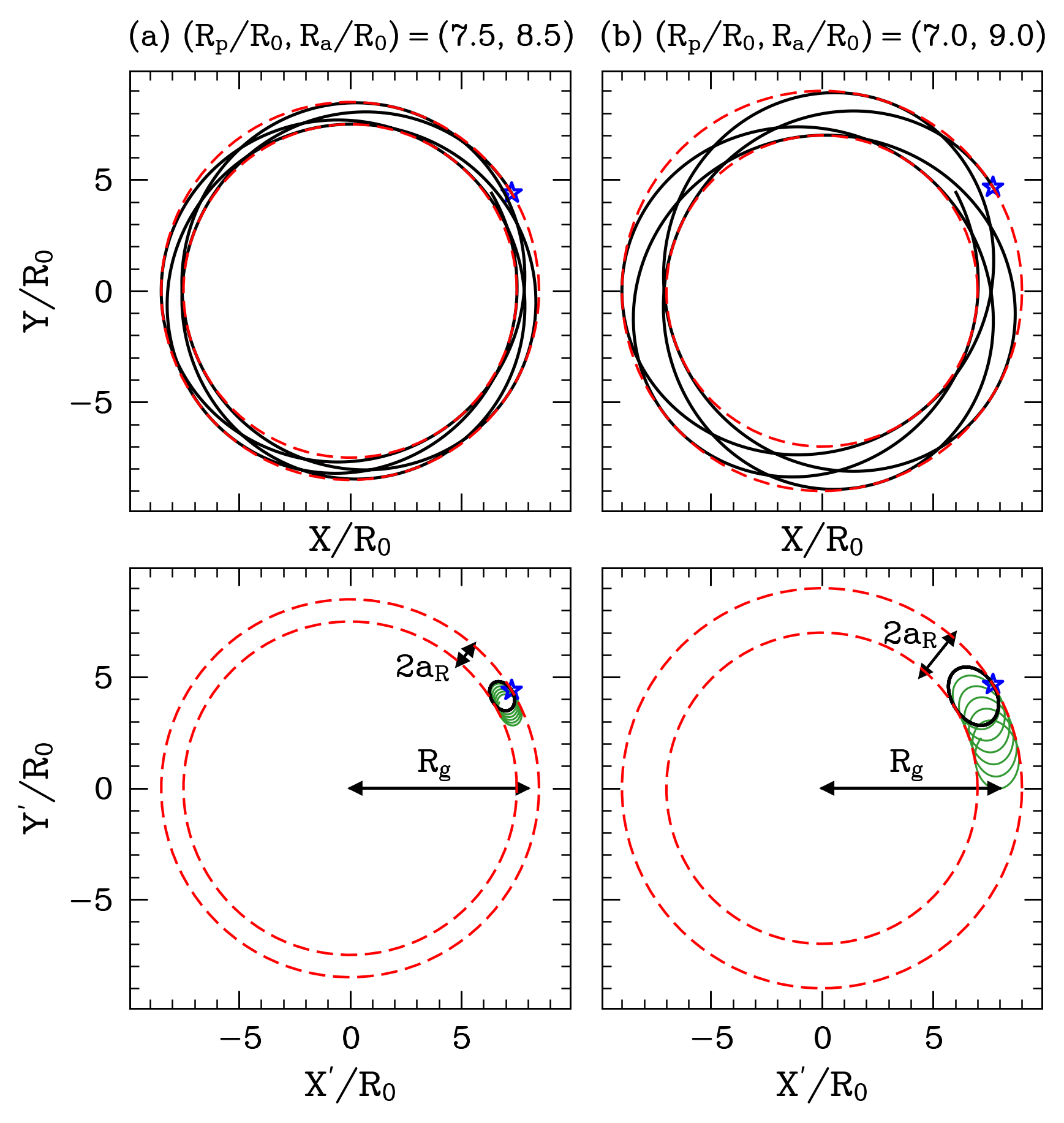}
    \caption{Two orbits in the logarithmic potential.
    The peri/apocenter distances $(\Rp, \Ra)$ for each orbit are shown with red dashed circles, and the initial location is shown with a small blue star. Full details of the orbital parameters are given in Table \ref{table:three_orbits}.
    The top row  shows the orbits as viewed in an inertial $(X,Y)$ frame. The bottom row shows the same orbits in the frame rotating azimuthally \red{with frequency} $\Omega_\varphi$ (equation \eqref{eqn:azi_freq}).
    In green we show what the orbits would look like if we had ignored the \red{Dehnen drift correction} in \eqref{eqn:azi_freq} and just rotated at $\Omega$.}
    \label{fig:three_orbits_rotating}
\end{figure}

As an example, we integrated two test-particle orbits in the logarithmic potential 
\begin{equation}
    \widetilde{\phi}(R) = V_0^2 \ln  \frac{R}{R_0},
    \label{eqn:log_halo}
\end{equation}
with $R_0$ an arbitrary length scale. This corresponds to a galaxy with a flat rotation curve $V(R)=V_0=$ const. (We will employ this potential for examples throughout the paper; 
to convert to physical units, a sensible choice is $R_0 = 1$\,kpc and $V_0 = 220\mbox{ km s}^{-1}$.)
In Figure \ref{fig:three_orbits_rotating} we plot the trajectories of orbits which have the same `semimajor axis' $\mathrm{sma} \equiv (\Rp+\Ra)/2 = 8R_0$, where $(\Rp, \Ra)$
are the pericenter/apocenter of the orbit, but different `eccentricity' $\mathrm{ecc} \equiv (\Ra-\Rp)/(\Ra+\Rp)$.
Each orbit is integrated for roughly four azimuthal periods, i.e., from time $t=0$ to $t=200\,T_0$, where $T_0\equiv R_0/V_0$ is the natural time unit. Further properties of these orbits are summarized in Table \ref{table:three_orbits}; in particular, their values of $\epsilon_R = (J_R/J_\varphi)^{1/2}$ are $0.053$ and $0.11$ respectively. 

The top row of Figure \ref{fig:three_orbits_rotating} shows these two 
orbits (a) and (b) as viewed in an inertial frame, while the bottom row shows them in a frame rotating at the \red{(Dehnen-corrected) azimuthal frequency $\Omega_\varphi$ (equation \eqref{eqn:azi_freq}).
In this frame the orbits form near-perfect ellipses. Meanwhile the green lines in Figure \ref{fig:three_orbits_rotating} show the what the orbit would look like if we instead ignored the drift correction and rotated at frequency $\Omega$. A systematic error clearly accumulates in this approximation, especially at larger epicyclic amplitudes.}
%and the larger is the 

%error is larger  can prove  over time will turn outto be important in what follows.}
%While asymmetric drift affects the visual representation of orbits such as those in Figure \ref{fig:three_orbits_rotating}, we find that its dynamical impact is minor, so we mostly ignore it from here on (but see \S\ref{sec:one_transient_long}).}.

\medskip

%%%%%%%%%%%%%%%%%%%%%%%%%%%%%%%%%%%%%%%%
%%%%%%%%%%%%%%%%%%%%%%%%%%%%%%%%%%%%%%%%
\section{Potential fluctuations felt by epicyclic orbits}
\label{sec:Potential_Section}
%%%%%%%%%%%%%%%%%%%%%%%%%%%%%%%%%%%%%%%%
%%%%%%%%%%%%%%%%%%%%%%%%%%%%%%%%%%%%%%%%

In this section we introduce the main conceptual advance of our theory, which is to classify potential fluctuations by wavelength (\S\ref{sec:Wavelength_Regimes}) and then to find approximate expressions for the Fourier transforms of these fluctuations in each regime (\S\ref{sec:deltaphi_Expressions}).
We also test our expressions numerically (\S\ref{sec:Potential_Numerical}).

\red{We emphasize that throughout this paper we are assuming the potential fluctuation $\delta\phi$ is given,
and we do not seek to calculate it self-consistently through the Poisson equation. Nevertheless, the expressions we develop here will turn out to be exactly what we need for constructing a self-consistent theory, which is the focus of Paper II.}

%%%%%%%%%%%%%%%%%%%%%%%%%%%%%%%%%%%%%%%%
%%%%%%%%%%%%%%%%%%%%%%%%%%%%%%%%%%%%%%%%
\subsection{Classifying fluctuations by wavelength}
\label{sec:Wavelength_Regimes}
%%%%%%%%%%%%%%%%%%%%%%%%%%%%%%%%%%%%%%%%
%%%%%%%%%%%%%%%%%%%%%%%%%%%%%%%%%%%%%%%%

\red{Consider an arbitrary potential fluctuation $\delta \phi(\br)$ at a fixed time $t$.
Since $\varphi$ is a periodic coordinate we are always free to write 
\begin{align}
    \delta \phi(\br) &= \sum_{n_\varphi=-\infty}^\infty \delta \phi_{n_\varphi}(R) \me^{i n_\varphi \varphi},
\end{align}
where
\begin{align}
    \delta \phi_{n_\varphi}(R) \equiv \int_0^{2\pi} \frac{\md \varphi}{2\pi} \, \delta \phi(\varphi, R)\me^{-in_\varphi\varphi},
    \label{eqn:delta_phi_azimuthal_fourier}
    \end{align}
    is the $n_\varphi$th azimuthal Fourier component (note this integral is over $\varphi$, \textit{not} $\theta_\varphi$).
%for some real amplitudes $A_m(R)$ and real shape functions $B_m(R)$ with the properties $A_m = A_{-m}$, $A_0=0$,  and $B_m = -B_{-m}$.
%Moreover, any (differentiable) function $B_m(R)$ can be expressed as ${\int_0^R \md R'\,k_m(R')}$ where $k_m(R)= \md B_m/\md R$, meaning that without loss of generality, 
%where
%\begin{align}
%\delta \phi_m(R) &= A_m(R) \me^{i \int_0^R \md R'\,k_m(R')}.
%\end{align}}
%\red{Thus to specify a potential fluctuation we need only specify the real functions $A_m$ %and $k_m$.}

The key quantity in our theory is the local \textit{wavevector} of this fluctuation,
\begin{equation}
 \bm{k} = (k_\varphi, k_R),
 \end{equation} 
 where for a fixed $n_\varphi$ and guiding radius $R_\mathrm{g}$, we define the (real) azimuthal wavenumber
\begin{equation}
    k_\varphi \equiv \frac{n_\varphi}{R_\mathrm{g}},
    \label{eqn:azimuthal_wavenumber}
\end{equation}
and the (generally complex) radial wavenumber
\begin{equation}
    k_R  \equiv \left( \frac{-i}{\delta\phi_{n_\varphi}}\frac{\p \delta\phi_{n_\varphi}}{\p R}\right)_{R=R_\mathrm{g}}.
    %= \frac{-i}{A_m} \frac{\md  A_m}{\md R} + k_m,
    \label{eqn:radial_wavenumber}
\end{equation}
Here, and for the remainder of the paper, the $n_\varphi$ and $J_\varphi$ dependence of the wavevector $\bm{k}$ is implicit. Hereafter we will denote the absolute magnitude of $\bk$ by 
\begin{equation}
    k \equiv \vert \bk \vert = \sqrt{\vert k_\varphi \vert^2 + \vert k_R\vert^2}.
    \label{eqn:absolute_wavenumber}
\end{equation}}

%The key quantity in our theory is the characteristic radial wavenumber \textit{of the $n_\varphi$th  Fourier component} of the guiding center potential perturbation $\delta \Phi_{n_\varphi}$:
%\begin{equation}
%    \label{eq:dimensionless_radial_potential_derivative}
%   k \equiv  \frac{1}{\delta\Phi_{n_\varphi}}\frac{\p \delta\Phi_{n_\varphi}}{\p R_\mathrm{g}}.
%\end{equation}

%Note that $k$ is in general complex (because $\delta \Phi_{n_\varphi}$ is in general complex). In particular, in the case that the radial part of the potential takes the form $\delta \Phi_{n_\varphi}(R) \propto \exp(ik_R R)$ for real $k_R$, we have 
%\begin{equation}
%    k=ik_R,
%    \label{eqn:k_WKB}
%\end{equation}
%so that $\vert k \vert = \vert k_R \vert$.

We can now define three regimes of potential perturbations, which will be crucial to the rest of our investigation. Namely, \textit{at a fixed} $\bJ$, we have:
\begin{itemize}
    \item \underline{Long wavelength regime}: 
    \red{\begin{equation} 
    k \sim R_\mathrm{g}^{-1}, \,\,\,\,\,\,\,\,\, \mathrm{i.e.} \,\,\,\,\,\,\,\, k  a_R \sim \epsilon_R,
    \label{eqn:def_long_wavelength_regime}
    \end{equation}
    which requires both a long radial wavelength $\vert k_R\vert  \lesssim R_\mathrm{g}$ and a small number of `arms' $\vert n_\varphi \vert \sim 1$.} 
    %Since $\vert n_\varphi \vert$ can never be smaller than unity, it is not possible to have fluctuation wavelengths longer than this.
    This regime lies outside the range of validity of the WKB approximation. Typically, only the potentials arising from the lowest-harmonic number, longest-wavelength perturbations like bars, very open spirals, and dwarf galaxies in the outskirts of the disk (like the Large Magellanic Cloud) will fall into this regime.
    \item \underline{Intermediate wavelength regime}:
    \red{\begin{equation} 
    k \sim a_R^{-1}, \,\,\,\,\,\,\,\,\, \mathrm{i.e.} \,\,\,\,\,\,\,\, k  a_R \sim 1.
        \label{eqn:def_intermediate_wavelength_regime}
    \end{equation}
    This requires either 
    
    (i) $\vert k_R\vert  \lesssim R_\mathrm{g}^{-1}$ and $\vert n_\varphi \vert \sim \epsilon_R^{-1}$, or
    
    (ii)
    $\vert k_R\vert  \sim a_R^{-1}$ and $\vert n_\varphi \vert \sim 1$, 
    or 
    
    (iii) $\vert k_R\vert  \sim a_R^{-1}$ and $\vert n_\varphi \vert \sim \epsilon_R^{-1}$.

    Many realistic potential fluctuations, especially spiral arms, fall into this regime. }
    %In this case, the potential fluctuations oscillate in space sufficiently rapidly for the WKB approximation to work well, but still slowly enough to be relatively constant on the scale of an individual star's epicyclic orbit.}
    \item \underline{Short wavelength regime}:
   \red{\begin{equation} 
    k \sim R_\mathrm{g} a_R^{-2}, \,\,\,\,\,\,\,\,\, \mathrm{i.e.} \,\,\,\,\,\,\,\, k  a_R \sim \epsilon_R^{-1}.
         \label{eqn:def_short_wavelength_regime}
    \end{equation}
    This will be true whenever we have short radial wavelengths $\vert k_R \vert \sim R_\mathrm{g} a_R^{-2}$, or when we have a large harmonic number $\vert n_\varphi \vert \sim \epsilon_R^{-2}$, or both of these.
    ISM gas clouds tend to fall into this regime.
At these wavelengths the WKB approximation tends to work very well, and stars typically feel a potential that oscillates significantly on the scale of their epicyclic orbit.

Note that it is possible in principle to have fluctuations that are of even shorter wavelength than  \eqref{eqn:def_short_wavelength_regime} (e.g., $ka_R\sim\epsilon_R^{-2}$), by having even higher radial wavenumber (e.g., $\vert k_R \vert \sim R_\mathrm{g}^2 a_R^{-3}$) or even higher azimuthal number (e.g., $\vert n_\varphi \vert \sim \epsilon_R^{-3}$). However, in reality there tends to be very little gravitational power on these scales \citep{modak2025characterizing}. Moreover, the angle-action approach we are following here turns out to be rather impractical in such cases, as we discuss in Appendix \ref{sec:app_short}. Thus we will assume throughout this paper that the shortest wavelengths worth considering satisfy the ordering \eqref{eqn:def_short_wavelength_regime}.}
\end{itemize}

The way the potential fluctuations couple to stellar orbits will depend strongly upon whether they belong in the long, intermediate or short wavelength regime.
\red{In light of this, it} is worth writing down an estimate for a typical value of \red{$k a$, 
where $a$ is the rms value of $a_R$ (equation \eqref{eqn:rms_epicyclic_amplitude}),
in terms of the wavelength $\lambda \equiv 2\pi/k$.}
Using the identities from \S\ref{sec:epicyclic}, we have:
\begin{align}
    k a & \simeq \sqrt{2} \, \pi \gamma \frac{R_\mathrm{g}}{ \lambda } \frac{\sigma}{V}
    \\
    &\simeq  4.57 \times \left(\frac{\gamma}{\sqrt{2}}\right)\left(\frac{R_\mathrm{g}}{8\,\mathrm{kpc}}\right)\left(\frac{ \lambda }{1\,\mathrm{kpc}}\right)^{-1} \nn \\
    &\,\,\,\,\,\,\,\,\,\,\,\,\,\,\,\times\left(\frac{\sigma}{20\,\mathrm{km\,s}^{-1}}\right)\left(\frac{V}{220\,\mathrm{km\,s}^{-1}}\right)^{-1}.
\end{align}
\begin{figure}
    \centering
    \includegraphics[width=0.49\textwidth]{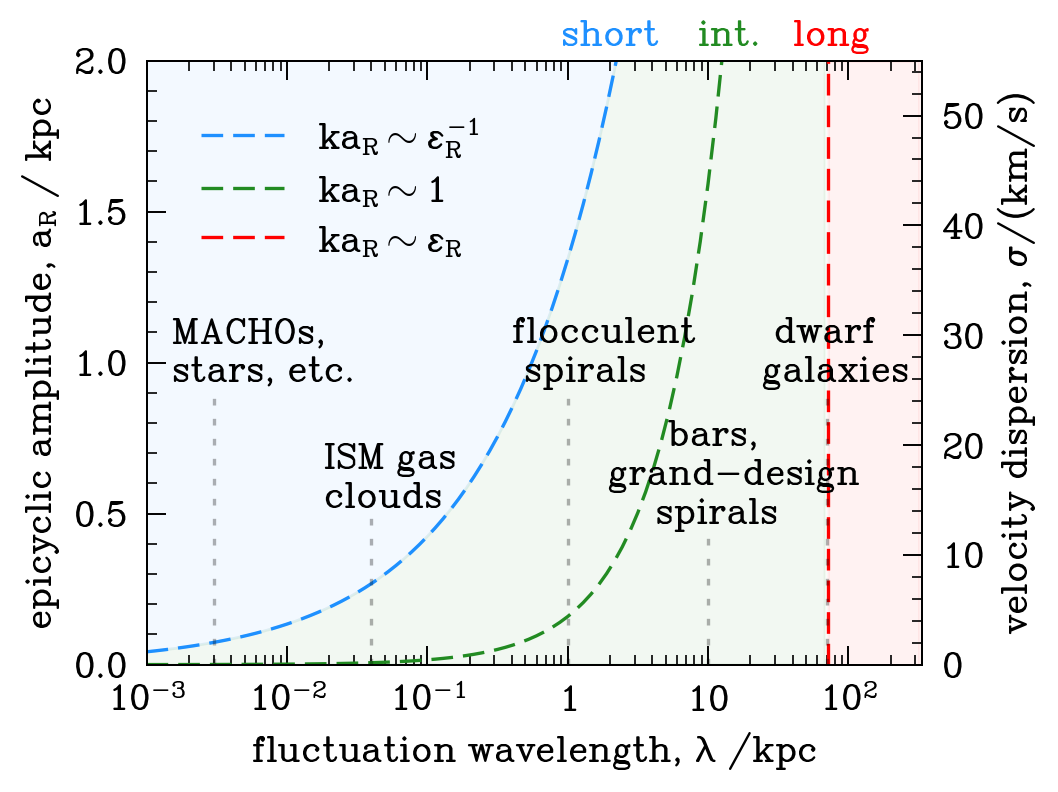}
    \caption{Diagram illustrating the short, intermediate and long wavelength regimes for various epicyclic amplitudes $a_R$,
    or equivalently velocity dispersion of the stellar population $\sigma$ (assuming $a_R=a$, see equations \eqref{eqn:rms_epicyclic_amplitude}-\eqref{eqn:rms_radial_velocity}). This plot assumes $R_\mathrm{g}=8\,$kpc, and a flat rotation curve ($\gamma=\sqrt{2}$) with circular velocity $V_0=220$\,km s$^{-1}$. We also indicate the typical wavelengths of potential fluctuations due to various astrophysical perturbations.}
    \label{fig:parameter_space}
\end{figure}
In Figure \ref{fig:parameter_space} we illustrate the parameter space of the three different wavelength regimes as a function of $ \lambda$ and epicyclic amplitude $a_R$ (or equivalently the velocity dispersion $\sigma$, calculated by assuming $a_R=a = \sqrt{2}\,\sigma/\kappa$). To make this plot we assumed $R_\mathrm{g}=8$ kpc, and a flat rotation curve (so $\gamma=\sqrt{2}$) with circular velocity $V_0=220\mbox{\,km\,s}^{-1}$, but the results are not particularly sensitive to these choices. We see that various types of perturbation --- ISM fluctuations, grand-design spirals, and so on --- can be categorized as short, intermediate or long wavelength, though their classification can depend on the velocity dispersion of the population of stars we are considering. 

\begin{figure*}
    \centering
    \includegraphics[width=0.85\textwidth]{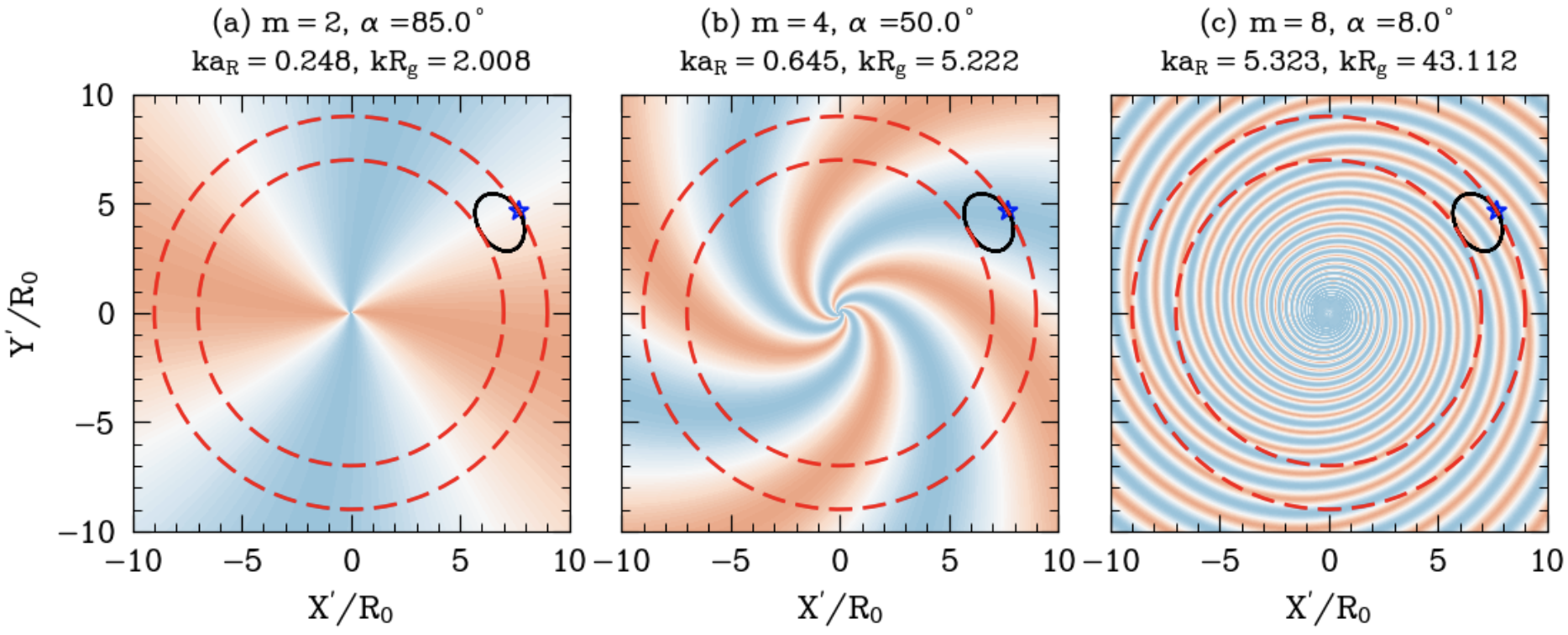}
    \caption{Orbit (b) from  Figure \ref{fig:three_orbits_rotating} \red{(which has $\epsilon_R=0.11$)}, superimposed on top of the contours of a logarithmic spiral potential \eqref{eqn:Phi_Spiral} with \red{$m$ arms and pitch angle $\alpha$.  The values of $m$ and $\alpha$} are chosen such that panels (a), (b) and (c) exhibit potential fluctuations characteristic of the long \red{($ka_R\sim\epsilon_R$)}, intermediate \red{($ka_R\sim 1$)} and short \red{($ka_R\sim\epsilon_R^{-1}$)} wavelength regimes (equations \eqref{eqn:def_long_wavelength_regime}-\eqref{eqn:def_short_wavelength_regime}) respectively.
    \red{We caution that the spirals are painted on this figure for illustrative purposes only; we have not chosen any pattern speed for them, and we do not mean to imply that the orbits shown are corotating with the spirals.}}
    \label{fig:three_pitches}
\end{figure*}

As a concrete example, in Figure \ref{fig:three_pitches} we reproduce the orbit from Figure \ref{fig:three_orbits_rotating}b \red{(which had $\epsilon_R=0.11$)}. In each panel of Figure \ref{fig:three_pitches} we superpose this epicyclic orbit on top of a logarithmic spiral potential\footnote{\red{Note that, strictly speaking, this potential is nonphysical in the sense that it cannot be generated by a finite surface density perturbation. A more physically-sensible choice would be the modified log-spiral potential discussed in \S2.6.3 of \cite{Binney2008-ou}, which has an extra prefactor $\propto R^{-1/2}$ compared to \eqref{eqn:Phi_Spiral}. In Paper II, where we are interested in self-consistent problems, we use modified log-spirals extensively.  Here, however, we are only interested in an illustrative example, so we stick with the pure log-spiral \eqref{eqn:Phi_Spiral} for simplicity.}}
\begin{align}
    &\delta \phi^\mathrm{ls}(\varphi, R, t) 
    \nn \\ 
    & \equiv -\eta \frac{ V_0^2}{2} \cos\left[m\cot\alpha\ln \frac{R}{R_0} + m(\varphi-\pattern t-\varphi_\mathrm{p})\right],
    \label{eqn:Phi_Spiral}
\end{align}
which corresponds to a spiral with dimensionless strength $\eta$, $m$ arms, pattern speed $\pattern$, pitch angle $\alpha$, and phase $\varphi_\mathrm{p}$. For this potential, all power is concentrated at azimuthal wavenumbers $n_\varphi = \pm m$. Using \eqref{eqn:azimuthal_wavenumber}-\eqref{eqn:absolute_wavenumber}, the \red{corresponding wavenumber $\bk = (k_\varphi, k_R)$ is given by
\begin{align}
     \bk &= \left( \frac{n_\varphi}{R_\mathrm{g}}, \frac{n_\varphi \cot \alpha}{R_\mathrm{g}} \right),
     \label{eqn:logspiral_wavenumber_components}
    \end{align}
    and its modulus $k = \vert \bk \vert$ is 
    %\frac{\vert m\,  \cot \alpha \vert }{R_\mathrm{g}}.
    \begin{align}
k=        \bigg\vert \frac{ n_\varphi }{R_\mathrm{g}\sin\alpha} \bigg\vert.
    \label{eqn:logspiral_wavenumber_modulus}
\end{align}}
In particular,  \red{$ k \to \vert n_\varphi\vert /R_\mathrm{g}$} as $\alpha \to 90^\circ$ (a very loosely wound spiral), and $ k \to \infty$ as $\alpha \to 0^\circ$ (a very tightly wound spiral). In Figure \ref{fig:three_pitches} we illustrate the cases \red{(a) $m=2, \, \alpha = 85^\circ$, (b) $m=4, \,\alpha = 50^\circ$ and (c) $m=8, \,\alpha = 8^\circ$. The resulting values of $ka_R$ and $k R_\mathrm{g}$ }  are given at the top of each panel. Panel (a) is characteristic of the long-wavelength regime \eqref{eqn:def_long_wavelength_regime}, panel (b) of the intermediate-wavelength regime \eqref{eqn:def_intermediate_wavelength_regime}, and panel (c) of the short-wavelength regime \eqref{eqn:def_short_wavelength_regime}.

%%%%%%%%%%%%%%%%%%%%%%%%%%%%%%%%%%%%%%%%
%%%%%%%%%%%%%%%%%%%%%%%%%%%%%%%%%%%%%%%%
\subsection{Expressions for potential fluctuations in asymptotic regimes}
\label{sec:deltaphi_Expressions}
%%%%%%%%%%%%%%%%%%%%%%%%%%%%%%%%%%%%%%%%
%%%%%%%%%%%%%%%%%%%%%%%%%%%%%%%%%%%%%%%%

Introducing the three wavelength regimes described above drastically simplifies the approximate expressions for the Fourier components of potential fluctuations $\delta \phi_{\bn}(\bJ)$. In Appendix \ref{sec:Potential_Theory} we derive explicit expressions for these functions \red{valid in the long and short wavelength regimes} up to corrections $\mathcal{O}(\epsilon_R^2)$. Here, we just gather the key results.
\\
\\
\underline{Long wavelengths}.  In this regime the Fourier components generically take the form
\red{\begin{align}
    \label{eq:deltaphi_cold_expansion}
    \delta \phi_{\bn}(\bJ) =  \delta \Phi_{n_\varphi}(J_\varphi)  \times  \begin{cases}
        1, & n_R=0, \\ 
        \frac{n_\varphi \gamma a_R}{2 R_\mathrm{g}} \left[ \pm 1 - \frac{ik_R R_\mathrm{g}}{n_\varphi \gamma}\right], & n_R = \pm 1,
        \\
        0, & \vert n_R\vert \geq 2.
    \end{cases} 
        \nn
    \\ \nn
    \\
\end{align}
} Thus, for $n_{R}=0$ we simply recover the identity \eqref{eq:Potential_Fluctuation_Fourier_Identify}; the potential fluctuation acts approximately as if the star were on a perfectly circular orbit. The terms arising from $n_R=\pm1$ are smaller by $\mathcal{O}(\epsilon_R)$; they are  proportional to $J_R^{1/2}$, and will ultimately be responsible for Lindblad resonances. Physically, the term involving $\pm 1$ stems from the azimuthal gradients in the potential, while the term proportional to \red{$k_R$} arises from radial gradients. Finally, for all other $\vert n_R\vert \geq 2$, long wavelength fluctuations \red{appear at $\mathcal{O}(\epsilon_R^2)$, so are} negligible to the accuracy we are working here.
\\
\\
\underline{Short wavelengths}. In this regime the potential fluctuation
$\delta \phi(\varphi, R)$ can be expanded as a sum of waves of the form 
\begin{equation}
    u_{n_\varphi}(k_R)\,\me^{i(n_\varphi\varphi + k_R R)}.
    \end{equation}
\red{The corresponding Fourier components are given in \eqref{eqn:deltaPhi_WKB_general}.
If we may ignore the Dehnen drift --- which is the case if $\vert n_\varphi\vert \sim 1$,  or if $\md \kappa/\md J_\varphi$ is small or zero as in the shearing sheet --- then this simplifies to the form \eqref{eqn:deltaPhi_WKB}.
Finally, for tightly wound waves with $\vert k_R / k_\varphi\vert \gtrsim \epsilon_R^{-2}$ (which is generically true in the short wavelength regime if $\vert n_\varphi \vert \sim 1$)
this reduces further to}
\begin{equation}
    \delta \phi_{\bn}(\bJ) = u_{n_\varphi}(k_R)\me^{i k_R R_\mathrm{g}}(-i \, \mathrm{sgn} \, k_R)^{n_R}J_{n_R}\left(\vert k_R\vert  a_R\right).
    \label{eqn:very_short_perturbations}
\end{equation}
However, the tight-winding approximation may not be \red{valid} if one is interested in, e.g., perturbations due to ISM density \red{fluctuations for which large-$\vert n_\varphi\vert$ structures are abundant \citep{meidt2023phangs}}.  For such structures it is better to rely on the more accurate \red{equation \eqref{eqn:deltaPhi_WKB} or even \eqref{eqn:deltaPhi_WKB_general}}.
\\
\\
\underline{Intermediate wavelengths}. %As we show in Appendix \ref{sec:Potential_Theory},
In the intermediate wavelength regime
there is technically no uniformly valid asymptotic expansion whose relative corrections are always $\mathcal{O}(\epsilon_R^2)$.  However, 
if the fluctuation is sufficiently tightly wound, \red{$\vert k_R / k_\varphi\vert \gg 1$, i.e., $\vert k\vert R_\mathrm{g} \gg \vert n_\varphi\vert$},
then one can naively extrapolate the result \eqref{eq:deltaphi_cold_expansion} from the long-wavelength regime to somewhat shorter wavelengths,
or extrapolate the result \eqref{eqn:very_short_perturbations} from the short-wavelength regime to somewhat longer wavelengths. In both cases one arrives at the same answer, 
namely 
\red{\begin{equation}
     \delta \phi_{\bn}(\bJ) =  \delta \Phi_{n_\varphi}(J_\varphi)\bigg[\delta_{n_R}^0
     -\frac{i}{2}k_Ra_R(\delta_{n_R}^1 + \delta_{n_R}^{-1}) \bigg] ,
\end{equation}}
where we identify $\delta \Phi_{n_\varphi}(J_\varphi) = u_{n_\varphi}(k_R)\me^{ik_R R_\mathrm{g}}$ (see equation \eqref{eqn:WKB}).
In this sense the intermediate regime provides a smooth interpolation between long and short wavelength regimes,
which suggests that \red{--- at least for $\vert n_\varphi \vert \sim 1$ ---} the approximate expressions \eqref{eq:deltaphi_cold_expansion} and \eqref{eqn:very_short_perturbations} will work well even beyond their strict range of validity.

%%%%%%%%%%%%%%%%%%%%%%%%%%%%%%%%%%%%%%%%%%%%%%%%%%%%%%%%%%%%
%%%%%%%%%%%%%%%%%%%%%%%%%%%%%%%%%%%%%%%%%%%%%%%%%%%%%%%%%%%%
%%%%%%%%%%%%%%%%%%%%%%%%%%%%%%%%%%%%%%%%%%%%%%%%%%%%%%%%%%%%

\subsection{Numerical examples}
\label{sec:Potential_Numerical}

%%%%%%%%%%%%%%%%%%%%
\begin{figure*}
    \centering
    \includegraphics[width=0.85\textwidth]{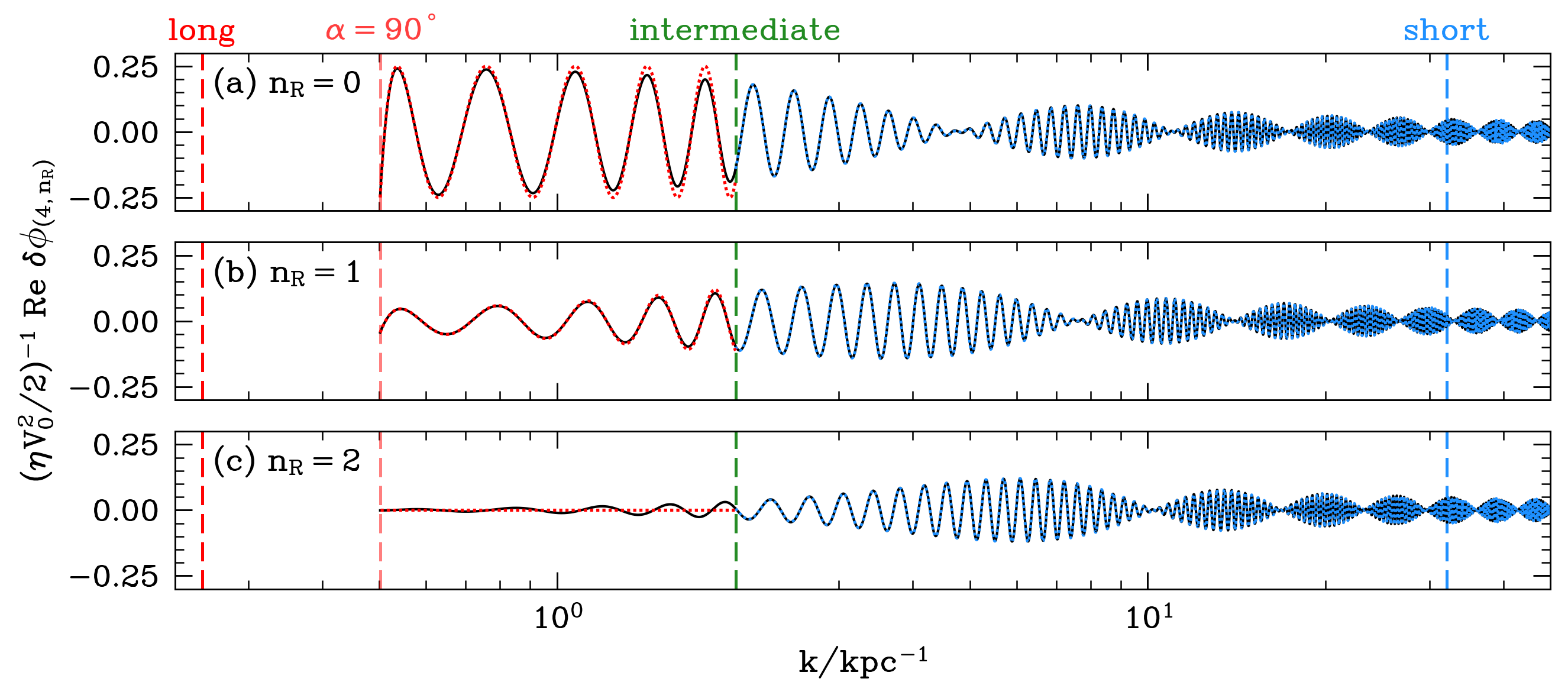}
    \caption{(Real part of the) Fourier components of the potential fluctuations $\delta \phi_{\bn}$ calculated for an $m=4$ logarithmic spiral potential perturbation \eqref{eqn:Phi_Spiral}, embedded in a logarithmic background potential \eqref{eqn:log_halo}, at the action-space location of the orbit shown in Figure \ref{fig:three_orbits_rotating}a (for which $\epsilon_R = 0.053$).
    From left to right we decrease the spiral's pitch angle and hence increase its  \red{wavenumber $ k$ (see equation \eqref{eqn:logspiral_wavenumber_modulus})}. Panels (a)-(c) show the result for $n_\varphi = 4$ and for $n_R = 0$, $1$ and $2$ respectively.
    The black line shows the result of evaluating the right-hand side of the expression \eqref{eq:potential_expansion_epicyclic} numerically, while the red and blue dotted lines show the asymptotic results \eqref{eq:deltaphi_cold_expansion} and \eqref{eqn:very_short_perturbations} respectively. See text for more details.
    }
    \label{fig:deltaphi_dependence_k}
\end{figure*}
\begin{figure*}
    \centering
    \includegraphics[width=0.85\textwidth]{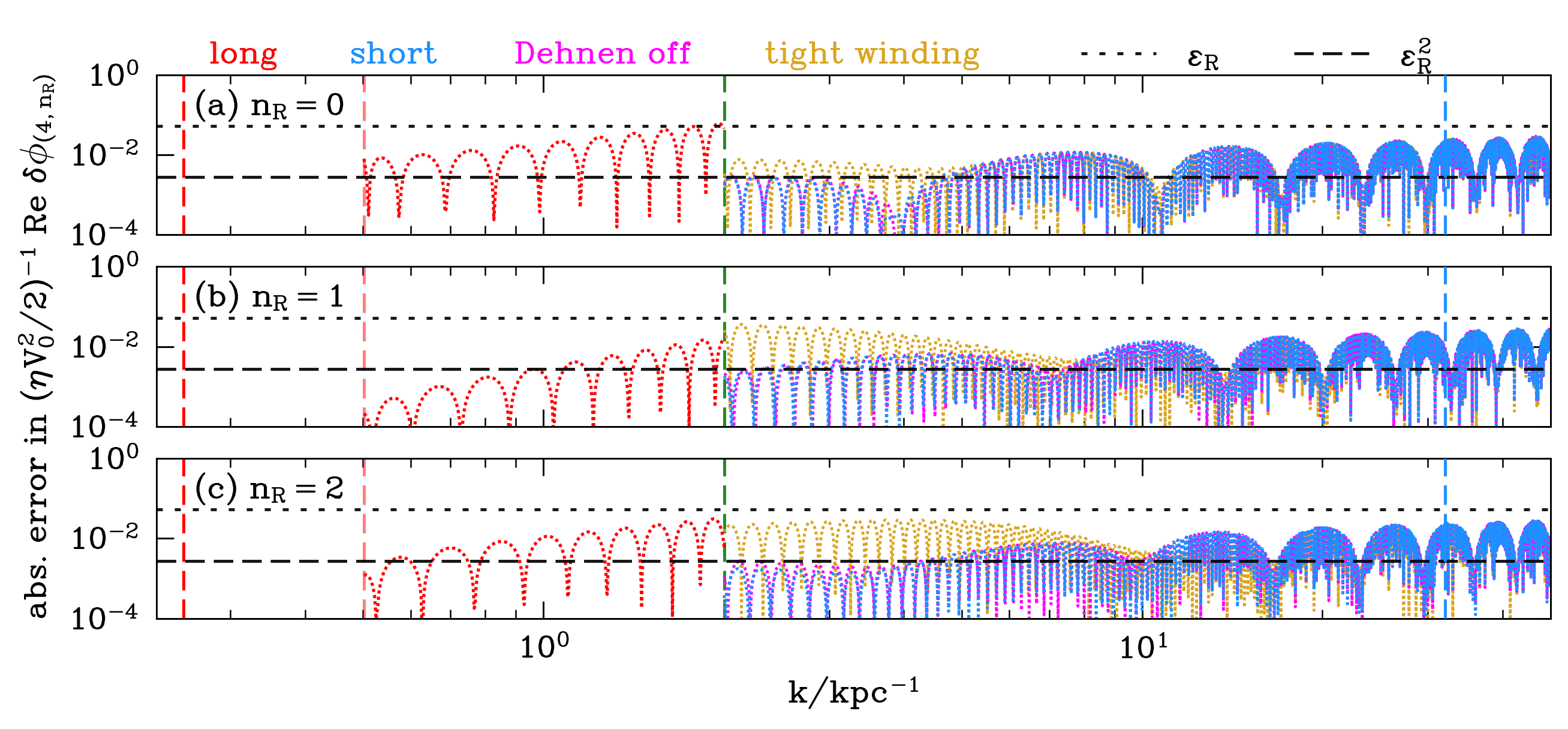}
    \caption{\red{As in Figure \ref{fig:deltaphi_dependence_k} except we plot the {absolute} difference between the numerical solution (the black line in Figure \ref{fig:deltaphi_dependence_k}) and the result of the long wavelength approximation (equation \eqref{eq:deltaphi_cold_expansion}, red), the short wavelength approximation (equation \eqref{eqn:deltaPhi_WKB_general}, blue), the same equation with the Dehnen terms dropped (equation \eqref{eqn:deltaPhi_WKB}, magenta),
    and the tight-winding approximation (equation \eqref{eqn:very_short_perturbations}, gold). We also indicate the magnitudes $\epsilon_R$ and $\epsilon_R^2$ with dotted and dashed black horizontal lines respectively.}
    }
    \label{fig:deltaphi_dependence_k_error}
\end{figure*}

%\begin{figure*}
%    \centering
%    \includegraphics[width=0.85\textwidth]{GK1_Re_deltaphi_Fourier_a_1.png}
%    \caption{As in Figure \ref{fig:deltaphi_dependence_k} except evaluated at the action-space location of the orbit shown in Figure \ref{fig:three_orbits_rotating}b.
%    Here $\epsilon_R = 0.11$ is about twice as large as in Figure \ref{fig:deltaphi_dependence_k}, leading to a slightly worse agreement between numerical and analytic results.}
%    \label{fig:deltaphi_dependence_k_b}
%\end{figure*}
%\begin{figure}
%    \centering
%    \includegraphics[width=0.49\textwidth]{GK1_Re_deltaphi_Fourier_a_1_WKB_Error.png}
%    \caption{\red{As in Figure \ref{fig:deltaphi_dependence_k_error} except for the orbit shown in Figure \ref{fig:three_orbits_rotating}b, for which $\epsilon_R = 0.11$.}
%    }
%    \label{fig:deltaphi_dependence_k_b_error}
%\end{figure}

\begin{figure}
    \centering
    \includegraphics[width=0.49\textwidth]{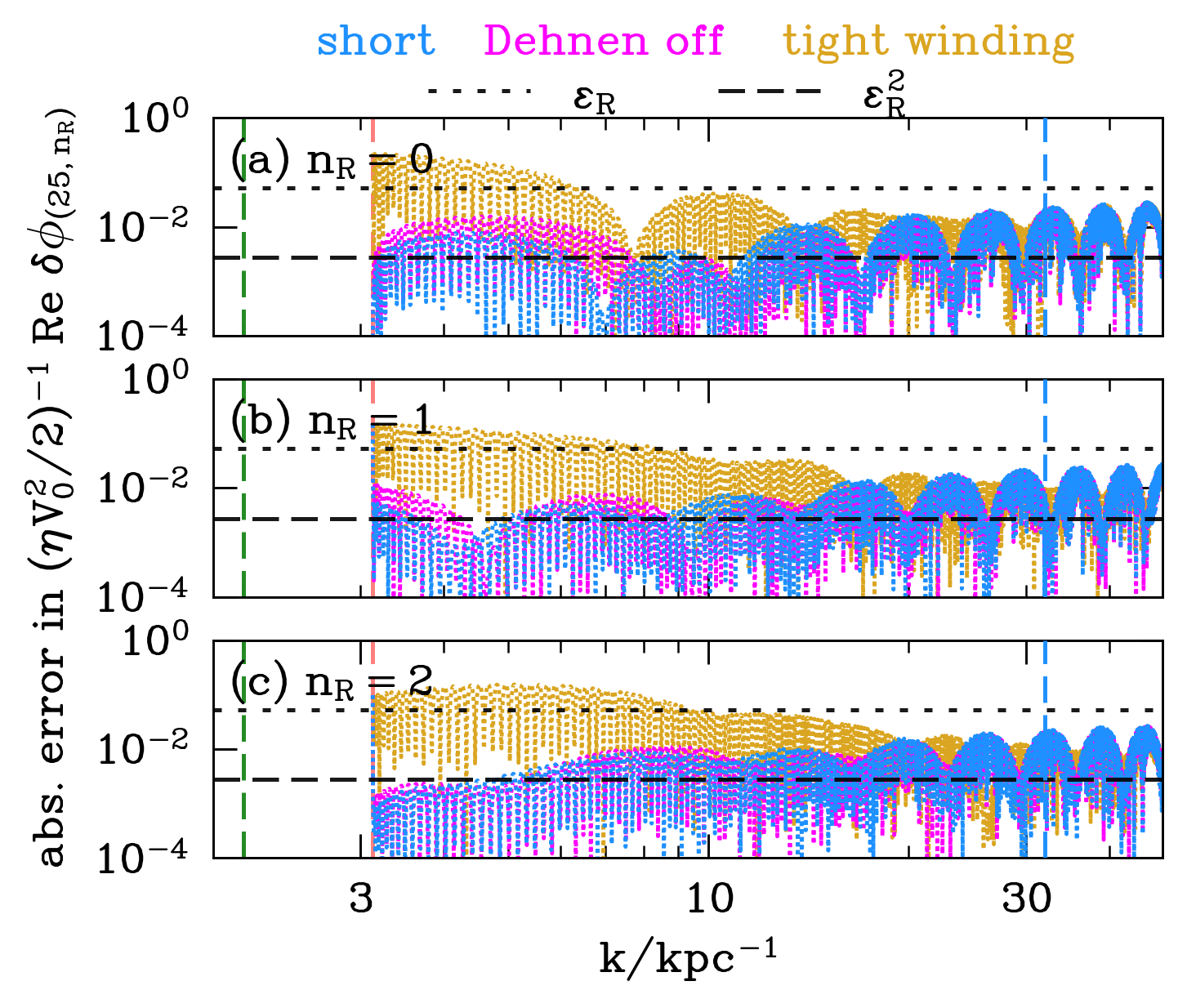}
    \caption{\red{As in Figure \ref{fig:deltaphi_dependence_k_error} except for an $m=25$ armed spiral perturbation. There are no long wavelength fluctuations in this case so we only show the intermediate/short wavelength regime.}
    }
    \label{fig:deltaphi_dependence_k_m25_error}
\end{figure}
In this subsection, we test the accuracy of our asymptotic expressions with some numerical examples.

First, we calculated $\delta \phi_{\bn}(\bJ)$ for the logarithmic spiral potential \eqref{eqn:Phi_Spiral} with $m=4$ arms, setting $t=0$ and $\varphi_\mathrm{p}=0$. To set up the mean field, we used the logarithmic potential \eqref{eqn:log_halo}, and we set $J_\varphi = 7.97 \,R_0V_0$ and $J_R = 0.022\,R_0V_0$, which are the actions of the orbit shown in Figure \ref{fig:three_orbits_rotating}a, and correspond to \red{$\epsilon_R= 0.053$ --- see Table \ref{table:three_orbits}. We computed $\delta \phi_{\bn}(\bJ)$ numerically using equation \eqref{eq:potential_expansion_epicyclic},  summing over $\ell,\ell' \in [-50, 50]$,
which is easily enough to ensure convergence.
We performed this calculation for different values of the pitch angle $\alpha$, i.e.,  for different radial wavenumbers $k_R$ (see equation \eqref{eqn:logspiral_wavenumber_components})}. In Figure \ref{fig:deltaphi_dependence_k} we plot with solid black lines the resulting values of Re $\delta \phi_{\bn}(\bJ)$ (normalized by $\eta V_0^2/2$) as a function of \red{$ k $ (equation \eqref{eqn:logspiral_wavenumber_modulus})} for $n_\varphi = 4$ and for $n_R=0, 1, 2$. 
Note that we have chosen $R_0=1$ kpc for this plot, so that wavenumbers are expressed in kpc$^{-1}$, and \red{the} orbit in question has a guiding radius of approximately $8$\,kpc and a rms radial velocity of about $20\,\mbox{km\,s}^{-1}$.

With vertical dashed lines we show  the values of $k$ corresponding to the long, intermediate and short asymptotic wavelength regimes; in addition, the the pink vertical dashed line shows the minimum possible $k = m/R_\mathrm{g}$ corresponding to $\alpha=90^\circ$. For all $k$ smaller than $a_R^{-1}$ (i.e., to the left of the green `intermediate' boundary) we plot 
the long wavelength result \eqref{eq:deltaphi_cold_expansion}
with a red dotted line;  for $k$ larger than $a_R^{-1}$ (to the right of the green boundary) we plot 
the most general short wavelength result \eqref{eqn:deltaPhi_WKB_general} with a blue dotted line.
%and 
%in the intermediate regime we overplot both asymptotic results simultaneously.
\red{Overall, the agreement between our analytic formulae and the exact numerical result looks very good.}

\red{We would like to test the accuracy of our approximations quantitatively, as well as the accuracy of the simpler short-wavelength equation \eqref{eqn:deltaPhi_WKB} in which the Dehnen terms are dropped, and/or its tightly wound cousin \eqref{eqn:very_short_perturbations}. To this end we present Figure \ref{fig:deltaphi_dependence_k_error}, in which we plot the \textit{difference} between the numerical solution (the black line in Figure \ref{fig:deltaphi_dependence_k}) and the various approximate analytic results. As expected, our long-wavelength approximation exhibits errors $\mathcal{O}(\epsilon_R^2)$ at long wavelengths but these become $\mathcal{O}(\epsilon_R)$ as we approach intermediate wavelengths. Meanwhile, the three short-wavelength approximations robustly exhibit errors between $\sim \epsilon_R$ and $\sim \epsilon_R^2$. Switching off the Dehnen terms makes almost no difference (the blue and magenta curves coincide almost exactly), but the tight-winding approximation (gold) exhibits a significantly larger error than the others close to the intermediate wavelength regime.}

\red{There are a couple of special features of Figure \ref{fig:deltaphi_dependence_k_error} that are not anticipated by our formalism, but are instead due to our choice of logarithmic spirals as model potential perturbations.
The first special feature is that, fortunately, the general WKB approximation \eqref{eqn:deltaPhi_WKB_general} turns out to be very good at intermediate wavelengths, exhibiting errors $\mathcal{O}(\epsilon_R^2)$ rather than the expected $\mathcal{O}(\epsilon_R)$.
The second special feature is that, unfortunately, all three short wavelength approximations give errors that grow roughly linearly with $k$ as we approach the short wavelength regime, so the error there can be slightly \textit{higher} than it was in the intermediate wavelength regime.
Both of these behaviors are due to the fact that when we substitute \eqref{eqn:Phi_Spiral} into \eqref{eq:potential_expansion_epicyclic}, the resulting integrand contains terms like 
$\delta \phi_{n_\varphi}(R) \propto \me^{i k_R R_\mathrm{g} \ln (R_\mathrm{g}-a_R\cos\theta_R)}$, which in the WKB approach we approximate with
$\delta \phi_{n_\varphi}(R) \propto \me^{i k_R (R_\mathrm{g}-a_R\cos\theta_R)}$.
When $k_Ra_R$ is not too large (as in the intermediate wavelength regime), the logarithmic term looks linear, and so the WKB approximation is excellent, but as $k_Ra_R$ gets bigger (as in the short wavelength regime), the curvature of the logarithm matters and the WKB approximation exhibits errors proportional to powers of $k_R$. 
Overall, these special features work to our advantage, because logarithmic spirals are a good basis for many realistic intermediate and long-wavelength fluctuations \citep{Kalnajs1976-gg, Binney2008-ou} where the WKB approximation works excellently.  Meanwhile at short wavelengths we rarely have logarithmic-spiral-like perturbations anyway\footnote{For instance, the Milky Way's spirals have $m=4$ and $\alpha \approx 12^\circ$ \citep{Eilers2020-na}, corresponding to $k \approx 2.4$ kpc$^{-1}$ (the intermediate regime) at the Solar radius.}, and it is often better to work directly with a spectrum of waves $\propto \me^{i\bk\cdot\br}$ as basis functions rather than log-spirals (c.f. the discussion in Appendix \ref{sec:app_short}).}
% Short wavelength fluctuations are not strongly suppressed as $\vert n_R \vert$ increases, but this causes no trouble since they are very well described by the WKB approximation.

\red{Finally, we tested our approximations for different orbits and different numbers of spiral arms $m$. For $m\sim 1$, the qualitative features were essentially unchanged from Figures \ref{fig:deltaphi_dependence_k}-\ref{fig:deltaphi_dependence_k_error}, with slightly larger errors as $m$ was gradually increased. Again, ignoring the Dehnen drift (i.e., using equation \eqref{eqn:deltaPhi_WKB}) did not make a significant difference compared to using the general WKB expression \eqref{eqn:deltaPhi_WKB_general}.
However, when we made $m$ significantly larger, the picture changed qualitatively.
In Figure \ref{fig:deltaphi_dependence_k_m25_error} we show the error profile from the same calculation as in Figures \ref{fig:deltaphi_dependence_k}-\ref{fig:deltaphi_dependence_k_error} except for $m=25$ (showing only intermediate/short wavelengths, since there is never a long wavelength regime in this case). We see that the general WKB expression \eqref{eqn:deltaPhi_WKB_general} (blue) continues to perform well, but that the simpler equation \eqref{eqn:deltaPhi_WKB} (magenta) that ignores the drift gives a worse result (though the corresponding errors are still $\lesssim \epsilon_R$ at all $k$ shown here). Finally, the tight-winding approximation \eqref{eqn:very_short_perturbations} (gold) performs poorly in this case, as we would expect for such a large harmonic number.}

%Though we have focused on logarithmic spirals for our numerical verification, our results should generalize to other types of perturbations. We conclude \red{that the asymptotic results} we have derived are rather accurate,
%\red{provided one does not apply the simplifications \eqref{eqn:deltaPhi_WKB} or \eqref{eqn:very_short_perturbations} where they are not valid}. %Care is needed in the intermediate regime, especially for $n_R \neq 0$. 
%Of course, here we have considered only a low-order expansion (ignoring corrections $\mathcal{O}(\epsilon_R^2)$), but more accurate \red{numerical evaluations} could be used if so desired \citep{}.

%%%%%%%%%%%%%%%%%%%%%%%%%%%%%%%%%%%%%%%%%%%%%%%%%%%%%%%%%%%%%%%
%%%%%%%%%%%%%%%%%%%%%%%%%%%%%%%%%%%%%%%%%%%%%%%%%%%%%%%%%%%%%%%

\medskip

\section{Kinetic theory}
\label{sec:kinetic_theory}

\red{We now illustrate how the asymptotic expressions for potential fluctuations we developed in \S\ref{sec:Potential_Section} can be used to simplify kinetic calculations.} More precisely, we are interested in how the DF $f(\btheta, \bJ)$ responds to potential perturbations $\delta \phi$ that are assumed `small' in the sense that 
at all relevant $(\btheta, \bJ)$,
\begin{equation}
   \eta  \equiv \bigg\vert \frac{ \delta \phi}{h_0} \bigg\vert \ll 1,
   \label{eqn:def_varepsilon}
\end{equation}
where $h_0$ is the mean-field Hamiltonian (see equation \eqref{eq:h}), with magnitude $\vert h_0\vert \sim V^2/2$). 

Linear kinetic theory is concerned with relative changes to the distribution function $\delta f /f_0$ that are $\mathcal{O}(\eta)$. Next order $\mathcal{O}(\eta^2)$ effects can often be approached through {quasilinear kinetic theory}, which we discuss in \S\ref{sec:Single_Transient}. However, before we can do any perturbation theory we must acknowledge the fact that we
have now introduced two small parameters, namely 
$\epsilon \ll 1$ (equation \eqref{eqn:def_epsilon}) and $\eta \ll 1$ (equation \eqref{eqn:def_varepsilon}), but we have not yet described how to order them with respect to each other.

%%%%%%%%%%%%%%%%%%%%%%%%%%%%%%%%%%%%%%%%%%%%%%%%%%%%%%%%
%%%%%%%%%%%%%%%%%%%%%%%%%%%%%%%%%%%%%%%%%%%%%%%%%%%%%%%%
\subsection{Ordering small parameters}
\label{sec:Ordering_Small_Parameters}
%%%%%%%%%%%%%%%%%%%%%%%%%%%%%%%%%%%%%%%%%%%%%%%%%%%%%%%%
%%%%%%%%%%%%%%%%%%%%%%%%%%%%%%%%%%%%%%%%%%%%%%%%%%%%%%%%

The parameter $\epsilon$ measures the \red{`temperature'} of the mean-field orbits, while $\eta$ measures the \red{energy in} the potential fluctuations. 
If $\epsilon \sim \eta$, we can simply expand our equations \eqref{eq:Fluctuation_Evolution_1D}-\eqref{eqn:Curly_G} in a single parameter $\epsilon$. The more difficult regimes are (a) $\epsilon \ll \eta$ or (b) $\epsilon \gg \eta$.

First, we deal with regime (a). In this case we argue that fluctuations in the potential will \red{typically heat the stellar population up to $\epsilon \gtrsim \eta$ on a dynamical time or less}. To see this, suppose for simplicity that $\epsilon( t = 0) = 0$, and that $\delta \phi$ is some stochastic (and non-adiabatic) forcing with wavelength $\lambda$ and correlation time $\tau$ (as measured in the frame of a circular orbit). Now ask, what is the value of $\epsilon$ after one orbital period $t_\mathrm{dyn}$? There are 3 possible cases to consider depending on the strength and duration of the perturbation:
\begin{itemize}
    \item \underline{Case I: Strong, coherent forcing.} Suppose the acceleration $\sim|\delta\phi|/\lambda$ is sufficiently strong and/or $\tau$ is sufficiently long that the stars `fall down the hill' of the potential to its bottom. \red{(This is what happens when orbits are trapped in the potential well of a rigidly-rotating spiral, for instance).} In this case, they pick up a velocity dispersion $\sigma \sim |\delta\phi|^{1/2}$, and hence achieve $\epsilon(t_\mathrm{dyn}) \sim \sigma/V \sim \eta^{1/2} \gg \eta$.
    
    \item \underline{Case II: Weak, coherent forcing.} \red{Suppose the acceleration is weak, and that $\tau \gtrsim t_\mathrm{dyn}$.} Then after a correlation time $\tau$ the stars will simply have been kicked by an amount $|\Delta \bm{v}(\tau)| \sim |\delta\phi|\tau/\lambda$, and the value of $\epsilon$ will have changed by $\Delta\epsilon(\tau) \sim |\Delta \bm{v}(\tau)|/V \sim \eta V \tau/\lambda \sim \eta (R_\mathrm{g}/\lambda)(\tau/t_\mathrm{dyn})$. We can now see that $\epsilon(t_\mathrm{dyn}) \sim  \Delta\epsilon(t_\mathrm{dyn}) \sim \eta(R_\mathrm{g}/\lambda)$. Since $R_\mathrm{g}/\lambda \gtrsim 1$ for \red{most} realistic perturbations (see Figure \ref{fig:parameter_space}), we have $\epsilon(t_\mathrm{dyn}) \gtrsim \eta$.
    
    \item \underline{Case III: Weak, incoherent forcing.} \red{Suppose the acceleration is weak, and that $\tau \ll t_\mathrm{dyn}$.} 
    \red{Then as in Case II $\Delta\epsilon(\tau) \sim \eta (R_\mathrm{g}/\lambda)(\tau/t_\mathrm{dyn})$. But now}
    $\epsilon$ will undergo a random walk with step-size $\Delta\epsilon(\tau)$, and we can estimate 
    $\epsilon(t_\mathrm{dyn}) \sim \Delta\epsilon(\tau)\times(t_\mathrm{dyn}/\tau)^{1/2}\sim \eta (R_\mathrm{g}/\lambda) (\tau/t_\mathrm{dyn})^{1/2}$.  \red{In theory one could construct artificial perturbations  such that this is either large or small compared to $\eta$. Realistically, though, small $\tau$ tends to mean small $\lambda$. The most important examples are ISM density fluctuations, which have $(\tau/t_\mathrm{dyn})^{1/2} \sim 0.1-0.2$ and $R_\mathrm{g}/\lambda\sim 10$ \citep{modak2025characterizing}, again giving $\epsilon \sim \eta$.}
\end{itemize}

Second, we consider regime (b), i.e., fluctuations with $\eta \ll \epsilon$. In this case, strictly speaking
we ought to extend our epicyclic expansion (\S\ref{sec:epicyclic}) and the corresponding potential approximations (\S\ref{sec:Potential_Section}) to higher order in $\epsilon_R$ than we have done so far. 
Suppose for example  that $\epsilon \sim \eta^{1/2}$ and we were doing linear theory (i.e., working at $\mathcal{O}(\eta)$).  Then we ought to modify both the long (equation \eqref{eq:deltaphi_cold_expansion}) and short (equation \eqref{eqn:very_short_perturbations}) wavelength expansions to keep terms $\mathcal{O}(\epsilon_R^2)$.
Whether these corrections are important or not depends on the question at hand and the desired accuracy of the prediction. We give an example where $\mathcal{O}(\epsilon_R^2)$ corrections do make a (minor) difference in \S\ref{sec:Single_Transient}.

\begin{figure}
    \centering
    \includegraphics[width=0.45\textwidth]{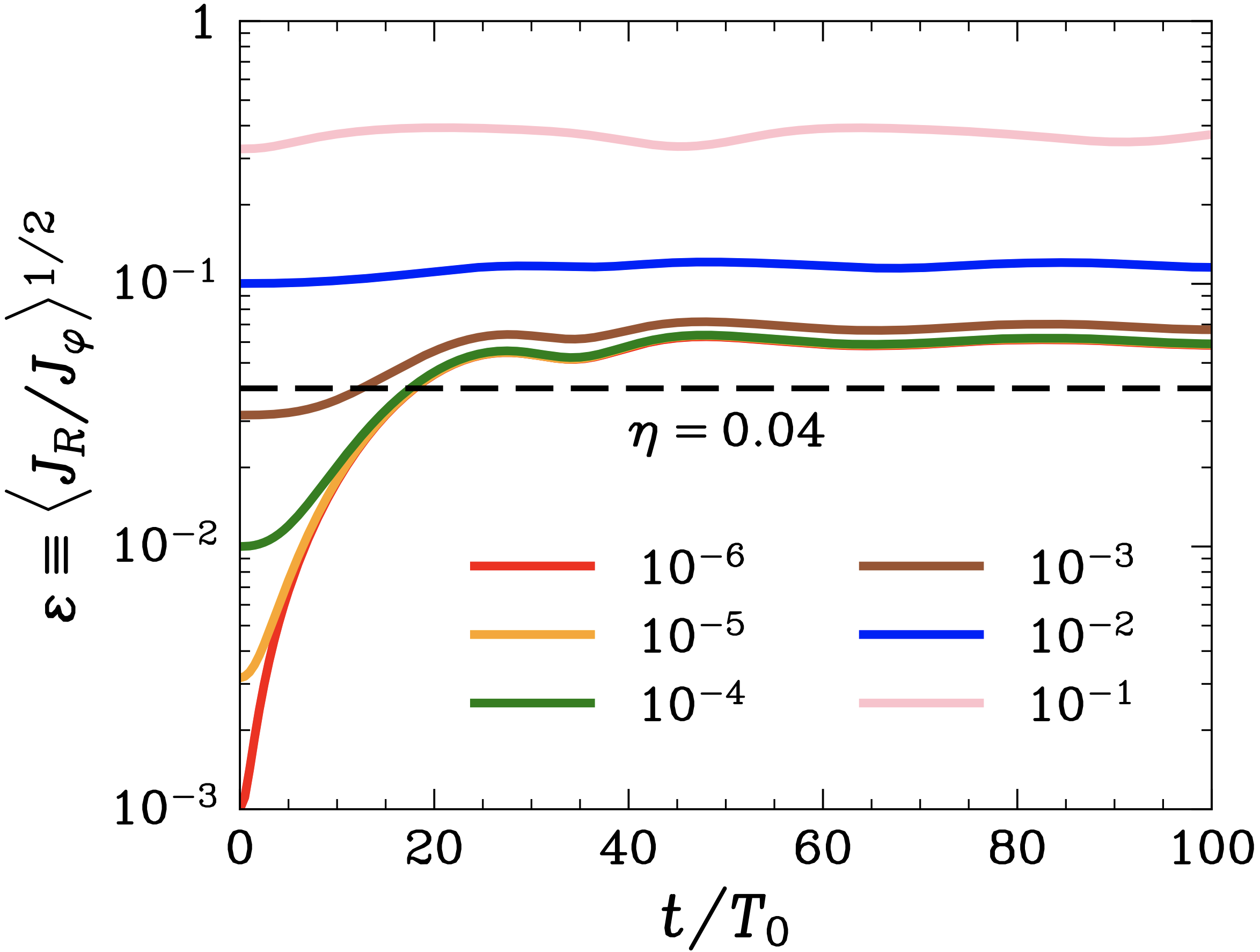}
    \caption{Evolution of $\epsilon \equiv \langle J_R / J_\varphi \rangle^{1/2}$ for ensembles of $10^4$ stars all with the same initial $J_\varphi$ but different initial $J_R$ when perturbed by a transient spiral (see text for details). The ensembles (shown in different colored curves) are labeled by their initial values of $J_R/J_\varphi = \epsilon(0)^2$, which vary from $10^{-6}$ to $10^{-1}$. Note that the azimuthal period at $J_\varphi = 8 (R_0V_0)$ is $T_\varphi \simeq 50T_0$). The black dashed line indicates the value $\eta = 0.04$; stars with $\epsilon(0) < \eta$ are rapidly heated so that $\epsilon(t_\mathrm{dyn}) > \eta$.}
    \label{fig:epsilon_t}
\end{figure}

To test these arguments, we simulated the motion of six ensembles of $10^4$ test stars in the logarithmic potential \eqref{eqn:log_halo}, subjected to a transient potential perturbation of the form
\begin{equation}
    \label{eqn:log_spiral_transient}
    \delta \phi(\varphi, R, t) = \me^{-(t-t_\mathrm{peak})^2/(2\tau^2)} \delta \phi^\mathrm{ls}(\varphi, R, t),
\end{equation}
where $\delta \phi^\mathrm{ls}$ is the logarithmic spiral potential \eqref{eqn:Phi_Spiral}. The potential \eqref{eqn:log_spiral_transient} grows and decays on a timescale $\sim \tau$, reaching maximum strength at $t_\mathrm{peak}$. We initialized all stars with angular momenta $J_\varphi = 8 \,R_0 V_0$ (in physical units, this might correspond to a star, like the Sun, whose guiding center is approximately $R_\mathrm{g} = 8$\,kpc) and random initial angles $\theta_\varphi \in (0,2\pi)$ and $\theta_R \in (0,2\pi)$. We altered the initial radial action from one ensemble to the next, such that $J_R/J_\varphi$ (and hence $\epsilon^2$, see equation \eqref{eqn:def_epsilon}) initially took the values $[10^{-6}, \, 10^{-5},\, 10^{-4},\, 10^{-3},\, 10^{-2}, 10^{-1}]$. 
We perturbed each ensemble with a trailing spiral of the form \eqref{eqn:log_spiral_transient}, with $m = 4$, pitch angle $\alpha = 50^\circ$ (matching the parameters of the `intermediate wavelength' regime example shown in panel (b) of Figure \ref{fig:three_pitches}), dimensionless amplitude $\eta = 0.04$, pattern speed $\Omega_\mathrm{p} = (8T_0)^{-1}$ (in physical units, this might correspond to a spiral with corotation radius $R_\mathrm{CR} = 8\,$kpc), initial phase $\varphi_\mathrm{p} = 0$, peak time $t_\mathrm{peak} = 8\pi T_0$ (corresponding to half an azimuthal period), and decay time $\tau = 4\pi T_0$ (corresponding to a quarter of an azimuthal period).

In Figure \ref{fig:epsilon_t} we plot the resulting evolution of $\epsilon(t)$ for each of these ensembles, shown with different colored curves, over just the first $\sim 100\,T_0$ of the simulation (roughly two azimuthal periods)\footnote{Strictly, this measured $\epsilon$ is not exactly equivalent to the $\epsilon$ we defined at fixed $J_\varphi$ in equation \eqref{eqn:def_epsilon}, 
but we checked that this subtlety has no effect on our conclusions.
}. To guide the eye, the black dashed line indicates the value of $\eta = 0.04$. We see that on this timescale, the $\epsilon$ value for the  blue and pink ensembles (those with $\epsilon(0) \gg \eta$, i.e., regime (b)) undergoes only a very modest growth. On the other hand, the red, yellow and green ensembles (those with $\epsilon(0) \ll \eta$, i.e., regime (a)) heat up so that $\epsilon \gtrsim \eta$ within one orbital period. Thereafter, these ensembles exhibit similar modest growth to the initially hotter ensembles.

This behavior turns out to be generic: we found very similar phenomenology when we varied the correlation time $\tau$ and the strength $\eta$ of the perturbation. We have also found the same behavior when integrating test particles in shearing boxes filled with ISM density fluctuations (\red{Modak et al, in prep.}).

In summary, either the system is born in the regime $\epsilon \sim \eta$, or (a) it is born with $\epsilon \ll \eta$  but transitions to $\epsilon \sim \eta$ after a time $\lesssim t_\mathrm{dyn}$, or (b) it is born with $\epsilon \gg \eta$ and \red{strictly we should extend the results of \S\ref{sec:Potential_Section} to higher order in $\epsilon$}. 
For the remainder of this paper we will assume that $\epsilon \sim \eta$ and expand our equations using a single small parameter $\epsilon \ll 1$.

\subsection{Quasilinear theory}
\label{sec:Single_Transient}

`Quasilinear theory' refers to any kinetic calculation in which an $\mathcal{O}(\eta^2)$ effect 
can be computed by multiplying two linear, $\mathcal{O}(\eta)$, fluctuations together.
Here we study the quasilinear transport of angular momentum due to a single transient potential perturbation of the form
\begin{align}
    \delta \phi(\varphi, R, t) &= \, \me^{-(t-t_\mathrm{peak})^2/(2\tau^2)}  \nn \\ &\,\,\,\,\,\,\,\times S(R)  \cos [m\left(\varphi - \pattern t + g(R)\right)] ,
    \label{eqn:Gaussian_Transient}
\end{align}
for some shape function $m g(R)$, where $m>0$.
Throughout this section we ignore any self-gravity of the perturbed part of the DF, i.e., we treat the stars as test particles, so the potential fluctuation is exactly \eqref{eqn:Gaussian_Transient}.
\red{This also means that there is no backreaction of the stellar distribution on to the external perturber, so the energy and angular momentum of the whole system is not conserved.}

The response of \textit{individual} stellar orbits to this kind of perturbation has been well-studied \citep{hamilton2024kinetic}.
In particular, a quasilinear description of the orbits is typically valid as long as one can ignore nonlinear trapping at resonances, \red{which occurs roughly on the timescale} (see, e.g., \S IV of \citealt{hamilton2024kinetic}):
\begin{align}
    t_\mathrm{lib} & \sim  1 \, \mathrm{Gyr} \times \bigg( \frac{m}{2} \bigg)^{-1}
\bigg( \frac{\eta}{0.01} \bigg)^{-1/2}
\bigg( \frac{t_\mathrm{dyn}}{200\, \mathrm{Myr}} \bigg),
\label{eqn:t_lib_CR}
\end{align}
\red{where $\vert S(R_\mathrm{CR})/ h_0 \vert \sim \eta$.}
%The pendulum formalism is not really applicable for near-circular orbits at Lindblad resonances, but the libration timescale for these is not too dissimilar from \eqref{eqn:t_lib_CR}.
The calculation we will perform ignores trapping, and so is valid for $\tau \ll t_\mathrm{lib}$.
It resembles the work of \cite{Carlberg1985-gf}, with two small but significant differences.
First, our assumption of a Gaussian envelope in \eqref{eqn:Gaussian_Transient} is different from their exponential growth and decay, which has the minor advantage that there is no  discontinuity in slope at $t=t_\mathrm{peak}$. Second, \cite{Carlberg1985-gf} focused on calculating the heating $\Delta\langle J_R\rangle$, whereas we are going to focus on the change in the marginalized angular momentum DF $\Delta F_0(J_\varphi)$.
An analogous calculation in the opposite limit where $\tau \gg t_\mathrm{lib}$ has been performed by \cite{sridhar2019renewal}.

\subsubsection{Angular-momentum transport}
\label{sec:quasilinear_response_to_transient}

We first write down a formal expression for the linear response of $f$ to the perturbation $\delta \phi$ by ignoring the right-hand side of \eqref{eq:Fluctuation_Evolution}  and integrating in time.
Assuming an unperturbed initial condition we get 
%, and assume $f_0$ is time-independent (since it changes only nonlinearly, see equation \eqref{eq:Axisymmetric_Evolution}):
\begin{align}
    \label{eq:linear_Vlasov_formal_solution}
    \delta f_{{\bm{n}}}(\bm{J}, t) =    i \int_0^t \md t' \,\bn \cdot \frac{\p f_0(\bJ, t')}{\p \bJ}  \me^{- i {\bm{n}}\cdot \bOm (t-t')} \delta\phi_{{\bm{n}}}(\bm{J}, t').
\end{align}
Plugging this in to \eqref{eqn:phi_flux_Fourier} gives
\begin{align}
\mathcal{F}_0  &=  - 2\pi \sum_{\bn} n_\varphi \int_0^\infty \md J_R \int_0^t \md t' \, \bn \cdot \frac{\p f_0(\bJ, t')}{\p \bJ} \nn
  \\
 & \,\,\,\,\,\,\,\,\,
 \times  
 \me^{- i {\bm{n}}\cdot \bOm (t-t')} \delta\phi_{{\bm{n}}}(\bm{J}, t')
\delta\phi^*_{\bn}(\bJ,t)   . 
\label{eqn:proto_flux}
\end{align}
Let us now specialize to perturbations of the form  \eqref{eqn:Gaussian_Transient}. 
The Fourier components of this are
\begin{align}
    \label{eq:deltaPhinJt}
    \delta \phi_{\bm{n}}(\bJ, t) = 
        \psi_{\bm{n}}(\bJ)\me^{-i n_\varphi \pattern t} \me^{-(t-t_\mathrm{peak})^2/(2\tau^2)},
\end{align}
where 
\begin{equation}
    \psi_{\bm{n}}(\bJ) = \frac{1}{4\pi} \int {\md \theta_R} \me^{-in_R\theta_R} S(R) \me^{ in_\varphi [\varphi - \theta_\varphi + g(R)]},
    \end{equation}
for $n_\varphi = \pm m$, and $\psi_{\bm{n}}(\bJ) =0$ otherwise \red{(note that both $R$ and the combination $\varphi-\theta_\varphi$ depend only on $\theta_R$ and $\bJ$, but not on $\theta_\varphi$.)}.
Plugging \eqref{eq:deltaPhinJt} into the right-hand side of \eqref{eqn:proto_flux},
inserting the resulting expression into \eqref{eq:Mean_Evolution_1D}, and using the epicyclic approximation, we get
\begin{align}
   \frac{\p F_0}{\p t} &= 2\pi \frac{\p}{\p J_\varphi} \sum_{n_\varphi = \pm m}\sum_{n_R}n_\varphi
   \int_0^\infty \md J_R \, \vert \psi_{\bn} (\bJ) \vert^2 
     \, \nn
    \\
    &\,\,\,\,\,\,\,\times \int_0^t  \md t' \bn \cdot \frac{\p f_0(\bJ, t')}{\p \bJ}
    \me^{-i\red{\omega_{\bn}(\bJ)} (t-t')}
     \, \nn
    \\
    &\,\,\,\,\,\,\,\times
    \me^{-(t-t_\mathrm{peak})^2/(2\tau^2)}
 \me^{-(t'-t_\mathrm{peak})^2/(2\tau^2)},
\label{eqn:transient_delta_F0_1}
\end{align}
where the frequency
\red{\begin{equation}
    \omega_{\bn}(\bJ) \equiv n_\varphi \Omega_\varphi + n_R \Omega_R - n_\varphi\pattern,
    \label{eqn:resonant_frequency_epicyclic}
\end{equation}
includes the Dehnen drift (see equations \eqref{eqn:azi_freq}-\eqref{eqn:rad_freq})}.
Assuming $t_\mathrm{peak}\gg\tau$ we can extend the lower limit of the $t'$ integration to $-\infty$. Moreover, for sufficiently short-lived perturbations we can ignore the slow time evolution of $f_0(\bJ,t')$ on the right-hand side of \eqref{eqn:transient_delta_F0_1} (since it changes only nonlinearly, see equation \eqref{eq:Axisymmetric_Evolution})
and integrate directly in time.
The result is $F_0(J_\varphi, t\to\infty) = F_0(J_\varphi, 0) + \Delta F_0(J_\varphi)$ where 
\begin{align}
    &\Delta F_0 = \frac{(8\pi^5)^{1/2}m}{\Gamma} \frac{\p}{\p J_\varphi} 
           \nn\\
        & \times\sum_{n_R} \int_0^\infty \md J_R \, \mathcal{R}_\Gamma(\omega_{mn_R})  \vert \psi_{mn_R} \vert^2  \left( m\frac{\p f_0}{\p J_\varphi} + n_R \frac{\p f_0}{\p J_R}\right)
  .
\label{eqn:transient_delta_F0_3}
\end{align}
Here $\Gamma \equiv (\sqrt{2}\tau)^{-1}$ is the width in frequency space associated with the finite lifetime $\tau$ of the perturbation, and \begin{equation}
    \mathcal{R}_\Gamma(\omega) \equiv \frac{1}{\sqrt{2\pi}\Gamma}\me^{-\omega^2/(2\Gamma^2)},
\end{equation} is a Gaussian response function.
Finally, one can calculate the change in total angular momentum $J_\mathrm{tot}(t) \equiv 2\pi \int \md J_\varphi F_0(J_\varphi, t)$ due to the transient by writing 
\begin{equation}
    \Delta J_\mathrm{tot} = 2\pi \int \md J_\varphi \, J_\varphi\,\Delta F_0,
    \label{eqn:Delta_J_tot_definition}
\end{equation}
and substituting \eqref{eqn:transient_delta_F0_3} into the right-hand side. 

\red{The expression \eqref{eqn:transient_delta_F0_3} simplifies dramatically if we specify the asymptotic wavelength regime of the fluctuations.
However, we must mention one subtlety, which is that here we are measuring $k_R$ relative to the root-mean-square epicyclic amplitude $a$, i.e., relative to the bulk of the stellar population,
rather than to the epicyclic amplitude $a_R(\bJ)$ associated with a particular action $\bJ$.
In this sense the names `short wavelength' and `intermediate wavelength'
do not have quite the same meanings as we defined back in \S\ref{sec:Wavelength_Regimes}.}

%It is crucial to consider the short and intermediate wavelength regimes together because, as is clear from Figure \ref{fig:parameter_space},
%a perturbation with, e.g., $\vert \lambda \vert \sim 1$ kpc will be of intermediate wavelength from the point of view of the least-eccentric (smallest $a_R$) orbits,
%and of short wavelength from the point of view of more eccentric orbits. The calculation we want to perform, namely evaluating the right-hand side of equation \eqref{eqn:transient_delta_F0_3}, requires us to integrate over all $a_R$ values, and so necessarily includes contributions from both regimes.

Next we use the results of \S\ref{sec:Potential_Section} to evaluate \eqref{eqn:transient_delta_F0_3} in 
the long/intermediate wavelength regime \red{(\S\ref{sec:one_transient_long}), and give a numerical example (\S\ref{sec:transient_example})}.  
%We relegate the corresponding short/intermediate wavelength calculation to Appendix \ref{sec:transient_short}.

\subsubsection{Long and intermediate wavelengths}
\label{sec:one_transient_long}

At long and intermediate wavelengths, equations \eqref{eq:deltaphi_cold_expansion} and \eqref{eq:epsilon_R}
tell us that we may approximate
\begin{equation}
   \vert \psi_{mn} \vert^2 = \vert \Psi_{m} \vert^2
       \bigg[\delta_{n_R}^0 + 
        \frac{J_R}{J^+_{m}}\delta_{n_R}^1 + \frac{J_R}{J^-_{m}}\delta_{n_R}^{-1}\bigg],
        \label{eqn:u_long}
\end{equation} 
where $\Psi_{m}(J_\varphi) \equiv \psi_{\bn}(J_\varphi, 0)$ (cf.\ equation \eqref{eq:Potential_Fluctuation_Fourier_Identify}) and \red{
\begin{equation}
    J^{\pm}_{m} (J_\varphi)\equiv \frac{4J_\varphi}{m^2 \gamma^3} 
    \bigg \vert 1 \mp  \frac{ik_R R_\mathrm{g}}{m \gamma} \bigg \vert^{-2},
    \label{eqn:J_scale}
\end{equation}}
is an angular-momentum scale. \red{We substitute this into 
equation \eqref{eqn:transient_delta_F0_3} and assume $m$ is sufficiently small that 
$\vert m \,\partial f_0/\p J_\varphi \vert \ll \vert \p f_0/\p J_R \vert$. The resulting integrals over $J_R$ and $J_\varphi$ are then easy to do numerically. In fact the $J_R$ integral can be done \textit{analytically} in the case where we ignore the Dehnen drift in $\omega_{mn_R}$, i.e., we set $\Omega_\varphi \to \Omega$ in \eqref{eqn:resonant_frequency_epicyclic}; in that case
}
equation \eqref{eqn:transient_delta_F0_3} becomes
\begin{align}
    \Delta F_0  = \frac{(8\pi^5)^{1/2}m}{\Gamma} \frac{\p}{\p J_\varphi}
    \left[ \Lambda_m \vert \Psi_m \vert^2
    \right],
    \label{eqn:Delta_F0_final}
\end{align}
where $\Lambda_m = \Lambda_m^0
    +
        \Lambda_m^+
        +
    \Lambda_m^-$ and 
\begin{equation}
    \Lambda_m^0= \frac{m}{2\pi}\mathcal{R}_\Gamma(\omega_{m,0}) \frac{\p F_0}{\p J_\varphi},
    \label{eqn:Lambda_CR}
\end{equation}
\begin{equation}
    \Lambda_m^\pm = \mp \frac{1}{2\pi} \mathcal{R}_\Gamma(\omega_{m,\pm 1}) \frac{F_0}{J^\pm_m}.
    \label{eqn:Lambda_ILROLR}
\end{equation}

\red{Whether we ignore the drift in \eqref{eqn:resonant_frequency_epicyclic} or not, for} small $\Gamma$ (large $\tau$) the response function $\mathcal{R}_\Gamma$ is sharply peaked around zero frequency.
In that case angular-momentum transport is localized near the {corotation resonance} $\omega_{m,0}=0$, 
and the {inner/outer Lindblad resonances} $\omega_{m,\pm 1}=0$.
Physically, assuming $\partial F_0/\partial J_\varphi < 0$, stars are transported across the corotation resonance from `left to right', i.e., from $J_\varphi < J_\varphi^\mathrm{CR}$ (where $\omega_{m, 0}$ is positive) to $J_\varphi > J_\varphi^\mathrm{CR}$ (where $\omega_{m, 0}$ is negative).  
Overall, the corotation resonance acts to flatten the DF of angular momenta in its own vicinity.
On the other hand, the inner (outer) Lindblad resonances always act to push stars to lower (higher) angular momenta, regardless of the gradient of $F_0$. This is what allows sharp features like 
grooves to build up in the angular-momentum DF.

\begin{figure}
    \centering
    \includegraphics[width=0.49\textwidth]{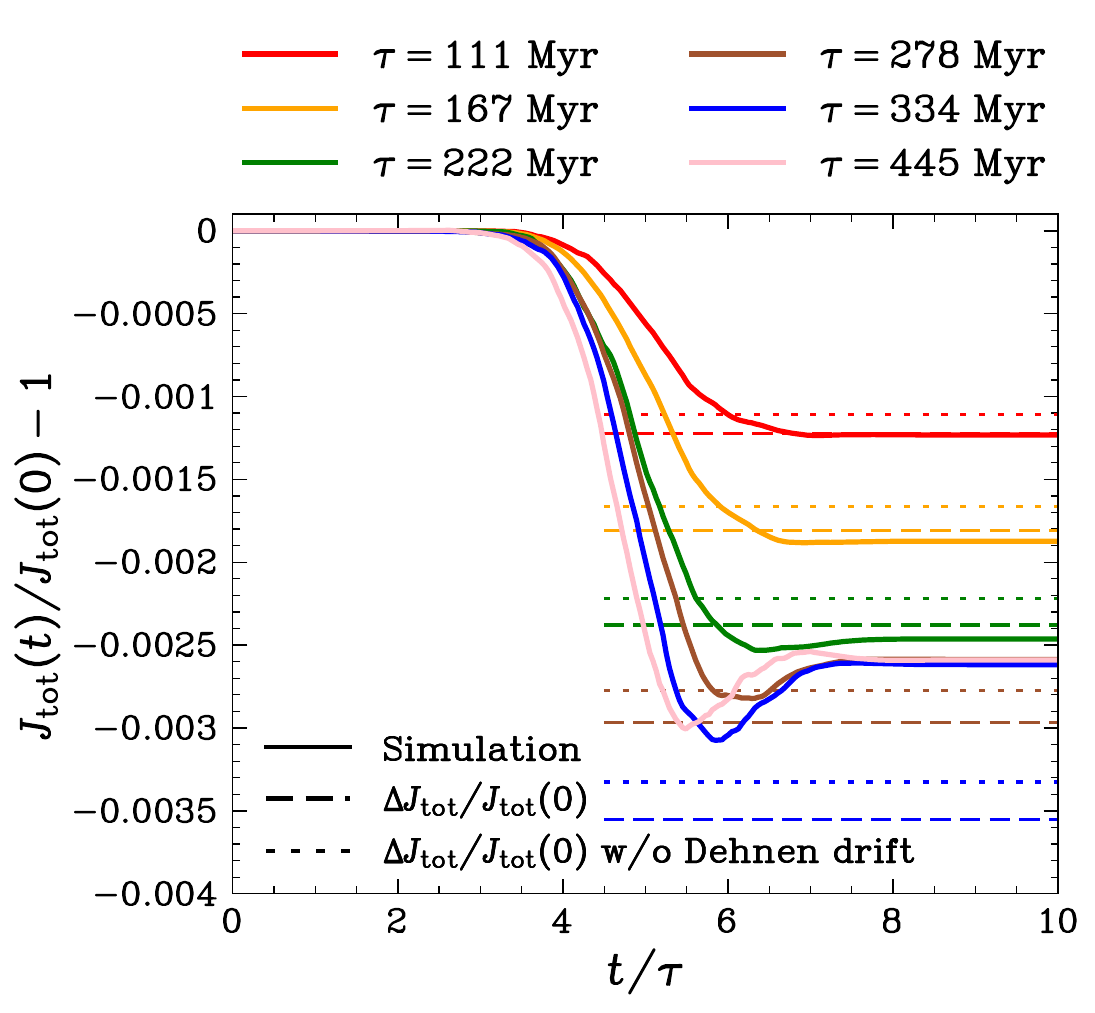}
    \caption{The change in total angular momentum $J_\mathrm{tot}$ of a disk of test particles due to an external transient perturbation with characteristic lifetime $\tau$.  For details of the setup, see \S\ref{sec:transient_example}. Solid lines show the results of $10^6$-particle simulations.  \red{Dashed (dotted) lines show the theoretical predictions of the total change, $\Delta J_\mathrm{tot}$, including (ignoring) the Dehnen drift correction to the orbital frequency $\Omega_\varphi$.}}
    \label{fig:DeltaJ_over_J}
\end{figure}

\subsubsection{\red{Numerical example}}
\label{sec:transient_example}

As \red{an} example, we consider ensembles of $10^6$ test stars orbiting in the logarithmic potential \eqref{eqn:log_halo}.
For the initial conditions of the stars
we choose random angles $\theta_\varphi \in [0, 2\pi)$ and $\theta_R \in [0, 2\pi)$, and random actions according to the DF $f_0(\mathbf{J}) \propto \exp(-J_\varphi/J_\varphi^\mathrm{s} )\exp(-J_R/\langle J_R\rangle)$, with parameters representative of the Milky Way: $J_\varphi^\mathrm{s} = 4\, \mathrm{kpc} \times V_0$ and $\langle J_R \rangle = 0.04 \,\mathrm{kpc} \times V_0$ (hence $\epsilon \sim  0.1$), where $V_0 = 220\mbox{\,km\,s}^{-1}$. For our perturbing potential $\delta \phi$ we take the form \eqref{eqn:log_spiral_transient}, with $m=4$ arms, strength $\eta=0.01$, pitch angle $\alpha = 90^\circ$, and pattern speed $\pattern$ chosen such that corotation is located at $R_\mathrm{CR}=8\,\mathrm{kpc}$.
We follow the orbits for a time $10 \tau$, where the characteristic growth/decay time $\tau$ is varied from one simulation to the next. We set $t_\mathrm{peak} = 5\tau$, such that the transient always peaks in amplitude half way through the simulation. The libration timescale \eqref{eqn:t_lib_CR} at peak amplitude in all of these simulations is $t_\mathrm{lib} \sim 0.5$ Gyr, comparable to the maximum $\tau = 445$ Myr.

In Figure \ref{fig:DeltaJ_over_J} we show the results of this experiment.
With solid curves we plot the simulation results for each decay time $\tau$ in various colors. \red{With dashed lines we show the theoretical prediction for the asymptotic change $\Delta J_\mathrm{tot}$, computed including the Dehnen drift as described below \eqref{eqn:J_scale}.
In addition, with dotted lines we show the corresponding result where we ignore the drift, i.e., plugging \eqref{eqn:Delta_F0_final} into \eqref{eqn:Delta_J_tot_definition}.}
We see that at the lowest $\tau$, \red{our theory does a very good job of predicting the final $J_\mathrm{tot}$ value, but we would have committed a $\sim 10\%$ error if we had ignored the drift, \textit{even though it contributes negligibly to the potential fluctuations in this regime} (equation \eqref{eq:deltaphi_cold_expansion}).
This is because of the extra ingredient, $\me^{-i\bn\cdot \bOm (t-t')}$, that is included in the linear perturbation calculation along with $\delta \phi_{\bn}$ (see equation \eqref{eq:linear_Vlasov_formal_solution}), which encodes the fact that stars with different epicyclic amplitudes gradually drift out of (azimuthal) phase with one another (see Figure \ref{fig:three_orbits_rotating}), so the resonance condition $\omega_{\bn}=0$ is a function not only of $J_\varphi$ but also $J_R$}.

As we increase $\tau$, \red{our theoretical result becomes} gradually \red{less accurate}. At the largest $\tau$ the theory over-predicts the magnitude of $\Delta J_\mathrm{tot}$, whereas the numerical result converges to a finite (negative) value. This numerical convergence, and the failure of the theory to predict it, occurs because of the time-dependence of $f_0(\bJ, t)$.
This can be induced either by (i) the quasilinear transport, which 
we ignored in writing down \eqref{eqn:transient_delta_F0_3}, 
or (ii) orbit trapping at resonances (for the largest $\tau$), which would correspond to a complete breakdown of quasilinear theory. 
While a detailed study is beyond the scope of this paper, 
both effects (i) and (ii) produce local flattening of $f_0(\bJ)$ at each resonance, meaning that eventually no more transport can occur there.

\medskip

\section{Conclusions}
\label{sec:Discussion}

In this paper we have introduced a scheme for approximating potential fluctuations in galactic disks, and shown how this renders tractable some important stellar-dynamical calculations.
Our findings can be summarized as follows.
\begin{itemize}
\item There are two key lengthscales involved in the description of in-plane orbital motion in a cool galactic disk: the guiding radius $R_\mathrm{g}$ and the typical epicyclic amplitude $a$ at that guiding radius.
For cool disks, $\epsilon \sim a/R_\mathrm{g} \ll 1$.
    \item Potential fluctuations in cool disks can be split into long, intermediate and short wavelength regimes relative to $R_\mathrm{g}$ and $a$. The Fourier-transformed potential fluctuation $\delta \phi_{\bn}(\bJ)$, which appears in all sophisticated kinetic theories, is drastically simplified in each asymptotic regime. We gave explicit asymptotic 
    expressions $\delta \phi_{\bn}(\bJ)$ that are accurate up to relative corrections $\mathcal{O}(\epsilon^2)$ \red{in the long and short wavelength regimes.
    At intermediate wavelengths the errors in our work are formally $\mathcal{O}(\epsilon)$, but, reassuringly, the long and short wavelength results connect smoothly here, rendering the theory valid over a larger range of scales than might be expected.}
    \red{We verified} these expressions numerically using the example of logarithmic spirals.
    %Moreover, our asymptotic expansions afford a smooth transition between these regimes.
 %   \item Our asymptotic potential expansion results in a simplification of linear response theory.  As an example we provided a 
 %   clean and physically transparent derivation of the groove instability.
    \item Our approach also renders tractable various \textit{quasilinear} expressions that in the past have often been evaluated numerically.  As an example, we calculated the quasilinear response of a cool stellar disk to a transient spiral perturbation. We verified the accuracy of this calculation via numerical simulation.
\end{itemize} 
%Admittedly, the latter example --- the quasilinear response to a transient spiral --- was already reasonably well-understood analytically by previous authors \citep{Carlberg1985-gf}.
%However, \red{one advantage of} our galactokinetic
%scheme is that it will allow us to investigate (semi-)analytically what has previously been studied only by brute-force numerics. 
%In an upcoming paper we will use it to address the crucial issue of radial heating and migration in the Milky Way.

Of course, our theory has its limitations. First, it \red{mostly} ignores corrections $\mathcal{O}(\epsilon^2)$ and higher, which can be on the order of several percent in the Galaxy. Higher order terms are also vital if one wishes to incorporate resonances with $\vert n_R\vert \geq 2$ into the long/intermediate wavelength theory. A natural extension would therefore be to \red{improve upon the epicyclic approximation} \citep{dehnen1999approximating} and \red{then perform the} associated potential expansion to the next order in $\epsilon$.
Second, our quasilinear theory ignores nonlinearly trapped orbits, \red{so a} next step would be to try to extend the theory to include multiple orbit families.
Third, we have ignored the self-consistent field $\delta \phi$ produced by the perturbation $\delta f$. 
It \red{is} desirable to \red{instead} have a self-consistent potential theory derived using the same asymptotic orderings as we used in \S\ref{sec:Potential_Section}. \red{This is the focus of Paper II \citep{GK2}, in which we present a unified linear theory of spiral structure in stellar disks.}

%Regardless of these shortcomings, if it does  indeed prove possible to develop a truly predictive kinetic theory of galactic disks --- one that can be meaningfully compared with observations and numerical simulations --- then galactokinetics is surely the place to start.

%% IMPORTANT! The old "\acknowledgment" command has be depreciated. It was
%% not robust enough to handle our new dual anonymous review requirements and
%% thus been replaced with the acknowledgment environment. If you try to 
%% compile with \acknowledgment you will get an error print to the screen
%% and in the compiled pdf.
%% 
%% Also note that the akcnowlodgment environment does not support long amounts of text. If you have a lot of people and institutions to acknowledge, do not use this command. Instead, create a new \section{Acknowledgments}.
\begin{acknowledgments}
We are grateful to the referee for a constructive report. We thank Eve Ostriker, James Binney and Walter Dehnen for helpful remarks.
C.H. is supported by the John N. Bahcall Fellowship Fund and the Sivian Fund at the Institute for Advanced Study. S.M. acknowledges support from the National Science Foundation Graduate Research Fellowship under Grant No. DGE-2039656.
This research was supported in part by grant no. NSF PHY-2309135 to the Kavli Institute for Theoretical Physics (KITP).
\end{acknowledgments}

%% To help institutions obtain information on the effectiveness of their 
%% telescopes the AAS Journals has created a group of keywords for telescope 
%% facilities.
%
%% Following the acknowledgments section, use the following syntax and the
%% \facility{} or \facilities{} macros to list the keywords of facilities used 
%% in the research for the paper.  Each keyword is check against the master 
%% list during copy editing.  Individual instruments can be provided in 
%% parentheses, after the keyword, but they are not verified.

\vspace{5mm}
%\facilities{HST(STIS), Swift(XRT and UVOT), AAVSO, CTIO:1.3m,
%CTIO:1.5m,CXO}

%% Similar to \facility{}, there is the optional \software command to allow 
%% authors a place to specify which programs were used during the creation of 
%% the manuscript. Authors should list each code and include either a
%% citation or url to the code inside ()s when available.

%\software{astropy \citep{2013A&A...558A..33A,2018AJ....156..123A},  
%          Cloudy \citep{2013RMxAA..49..137F}, 
%          Source Extractor \citep{1996A&AS..117..393B}
%          }

%% Appendix material should be preceded with a single \appendix command.
%% There should be a \section command for each appendix. Mark appendix
%% subsections with the same markup you use in the main body of the paper.

%% Each Appendix (indicated with \section) will be lettered A, B, C, etc.
%% The equation counter will reset when it encounters the \appendix
%% command and will number appendix equations (A1), (A2), etc. The
%% Figure and Table counter will not reset.

\appendix
\section{Explicit Fourier expansions}
\label{sec:Explicit_Fourier}

For ease of notation we drop all $\bJ$ and $t$ arguments in $\Omega_\varphi$, $\delta f$ and $\delta \phi$. Then we have
\begin{align}
    \label{eq:Piece_1}
    &\overline{ \left( \Omega_\varphi  \frac{\partial \delta f}{\partial \theta_\varphi}\right)} = 2\pi \sum_{n_\varphi}in_\varphi \me^{in_\varphi \theta_\varphi}  \int_0^\infty \md J_R \, \Omega_\varphi\, \delta f_{(n_\varphi, 0)},
    \end{align}

\begin{align}
    \label{eq:Piece_2}
    &\overline{ \left( \frac{\partial \delta \phi}{\partial \theta_\varphi} \frac{\partial f_0}{\partial J_\varphi}\right)} = 2\pi \sum_{n_\varphi}in_\varphi \me^{in_\varphi \theta_\varphi}
    \int_0^\infty \md J_R \, \frac{\p f_0}{\p J_\varphi}\, \delta \phi_{(n_\varphi, 0)},
\end{align}

\begin{align}
    \label{eq:Piece_3}
    &\mathcal{G} = 2\pi\sum_{n_\varphi, n'_\varphi} \me^{i(n_\varphi+n'_\varphi)\theta_\varphi}\sum_{n_R}\int_0^\infty \md J_R \,\delta f_{(n_\varphi, n_R)}\frac{\p \delta\phi_{(n_\varphi', -n_R)}}{\p J_\varphi},
\end{align}

\begin{align}
    \label{eq:Piece_4}
    &\mathcal{F} = - 2\pi \sum_{n_\varphi, n'_\varphi} i n'_\varphi \me^{i(n_\varphi+n'_\varphi)\theta_\varphi}\sum_{n_R}\int_0^\infty \md J_R \,\delta f_{(n_\varphi, n_R)} \delta\phi_{(n_\varphi', -n_R)}.
\end{align}

 \medskip
%%%%%%%%%%%%%%%%%%%%%%%%%%%%%%%%%%%%%%%%
%%%%%%%%%%%%%%%%%%%%%%%%%%%%%%%%%%%%%%%%
\section{Fourier components of potential fluctuations in asymptotic regimes}
\label{sec:Potential_Theory}

We need the Fourier coefficients $\delta \phi_{\bn}(\bJ)$ for a given potential $\delta\phi(\varphi, R)$.  Inverting \eqref{eq:Fourier_deltaphi}, the general formula is 
\begin{equation}
    \delta \phi_{\bn}(\bJ) = \frac{1}{(2\pi)^2} \int \md \btheta \,\delta \phi(\varphi(\btheta, \bJ), R(\btheta, \bJ)) \,\me^{-i\bn\cdot\btheta}.
    \label{eq:xxx}
\end{equation}
Plugging in the epicyclic formulae \eqref{eq:Epicycle_Azimuthal_Angle}--\eqref{eq:Epicycle_Radius}
we find that up to corrections $\mathcal{O}(\epsilon_R^2)$, \red{
\begin{align}
    \label{eq:potential_expansion_epicyclic}
    &\delta \phi_{\bn}(\bJ) = 
    \sum_{\ell=-\infty}^\infty \sum_{\ell'=-\infty}^\infty J_\ell\left( n_\varphi \gamma \frac{a_R}{R_\mathrm{g}} \right)
    J_{\ell'}\left( \frac{n_\varphi a_R^2}{4} \frac{\md \kappa}{\md J_\varphi} \right)
       \int_0^{2\pi} \frac{\md \theta_R}{2\pi}   \me^{-i ({n_R}-\ell-2\ell')\theta_R} 
 \delta \phi_{{n_\varphi}}\left(R_\mathrm{g} - a_R  \cos\theta_R\right),
\end{align}
where $\delta \phi_{n_\varphi}(R)$ is defined in \eqref{eqn:delta_phi_azimuthal_fourier}, and the $J_\ell$ and $J_{\ell'}$ are Bessel functions. In particular, the $J_{\ell'}$ Bessel functions encode the effect of the Dehnen drift, i.e., the third term in the epiyclic expansion \eqref{eq:Epicycle_Azimuthal_Angle}. Ignoring this drift would amount to replacing $J_{\ell'}(...) \to \delta_{\ell'}^0$ in the above expression.}

We get different explicit forms on the right-hand side of \red{\eqref{eq:potential_expansion_epicyclic}} depending on which asymptotic wavelength regime we are in.

\subsection{Long-wavelength regime}

Let us work at a fixed $\bJ$ and $\bn$. Supposing that we are in the long wavelength regime 
\eqref{eqn:def_long_wavelength_regime} we \red{certainly have $\vert k_R \vert a_R \lesssim \epsilon_R$ (where $k_R$ is defined in \eqref{eqn:radial_wavenumber})}. This means that we can expand
\red{\begin{align}
     \label{eq:potential_fluctutaion_radial_expansion}
     &\delta \phi_{{n_\varphi}}(R_\mathrm{g} - a_R  \cos\theta_R)
   = \delta \Phi_{n_\varphi}(J_\varphi)( 1 - i{k_R a_R}\cos\theta_R+ ...),
\end{align}}
where the discarded terms are \red{$\mathcal{O}\left(\vert k_R a_R\vert ^2\right) \lesssim \mathcal{O}(\epsilon_R^2)$} and $\delta\Phi_{n_\varphi}(J_\phi)$ is defined by equations (\ref{eq:Potential_Fluctuation_Guiding_Center}) and (\ref{eq:fourier_guidingcenter_phi}).
\red{We can now} plug the expansion \eqref{eq:potential_fluctutaion_radial_expansion} into \eqref{eq:potential_expansion_epicyclic} and perform the integral over $\theta_R$ \red{explicitly.
Also, we know that in the long wavelength regime $\vert n_\varphi\vert \sim 1$, 
meaning that unless there are atypically sharp gradients in the $\kappa(J_\varphi)$ profile, the argument of the $J_{\ell'}$ Bessel function is $\mathcal{O}(\epsilon_R^2)$. Thus we can employ
the small-argument expansion of the Bessel function for $\ell \geq 0$ and $x\geq 0$,
\begin{equation}
    J_\ell(x) \to \frac{1}{\ell !} \left(\frac{x}{2}\right)^\ell,
    \label{eqn:Bessel_small_argument}
\end{equation}
as well as the identities $J_{-\ell}(x)=J_{\ell}(-x) = (-1)^{\ell}J_{\ell}(x)$,
which tell us that up to corrections $\mathcal{O}(\epsilon_R^2)$ the Dehnen drift can be ignored in the long wavelength regime.}
Putting these pieces together the result is
\red{\begin{align}
   & \delta \phi_{\bn}(\bJ) =  \delta \Phi_{n_\varphi}(J_\varphi)\bigg\{ 
    J_{n_R}\left( n_\varphi\gamma\frac{a_R}{R_\mathrm{g}}\right)
    - \frac{i}{2} k_R a_R \left[J_{n_R-1}\left( n_\varphi\gamma\frac{a_R}{R_\mathrm{g}}\right)
    + J_{n_R+1}\left( n_\varphi\gamma\frac{a_R}{R_\mathrm{g}}\right)\right]
    \bigg\}.
    \label{eqn:proto_long_wavelength_fluctuation}
    \end{align}}
\red{Finally, the arguments of the Bessel functions are $\sim \epsilon_R$ so we can again use \eqref{eqn:Bessel_small_argument}.}
\red{With this, up to corrections $\mathcal{O}(\epsilon_R^2)$ we} find that at long wavelengths the potential fluctuations are given by \eqref{eq:deltaphi_cold_expansion}.

\subsection{Short-wavelength regime}
\label{sec:app_short}

Suppose instead that we are in the short wavelength regime \eqref{eqn:def_short_wavelength_regime}, 
so that \red{$kR_\mathrm{g} \sim \epsilon_R^{-2}$.} 
\red{Then we expect that we may use a WKB approximation,}
meaning that we may expand an arbitrary potential fluctuation as a sum of sinusoidal waves.
In particular we can express the \textit{azimuthal} Fourier components of the potential fluctuation  as  
\begin{equation}
    \delta \phi_{n_\varphi}(R) = u_{n_\varphi}(k_R)\,\me^{ik_R R},
    \label{eqn:WKB}
\end{equation}
where $u_{n_\varphi}(k_R)$ and $k_R$ depend only weakly  on $J_\varphi$.
\red{Relative corrections to this approximation should} be of order 
\red{$\mathcal{O}\left(\vert k  R_\mathrm{g}\vert^{-1}\right) \sim \mathcal{O}(\epsilon_R^2)$}, negligible to the accuracy we are working here.

\red{Note that we expect this WKB approximation to hold even in the limit of small $\vert k_R \vert$ provided $\vert n_\varphi \vert$ is sufficiently large --- i.e., even if the corresponding density perturbation spreads out radially from the galactic center like the spokes of a bicycle wheel \citep{zengyuan1989uniformly}. The reason for this is that the gradient of the fluctuation is much stronger perpendicular to a `spoke' than parallel to it, and so $\delta \phi_{\bn}$ is determined locally in radius even if the radial wavelength is formally very long. In reality, of course, galactic structures do not look like radial spokes; rather, the high-$\vert n_\varphi \vert$ perturbations we have in mind are  gaseous ISM fluctuations like those  in the PHANGS galaxies \citep{meidt2023phangs}.
These may well be oriented radially (i.e., have a small $\vert k_R \vert$) in some  small patch of the disk, but will typically be oriented differently in adjacent patches. Strictly speaking, the minimum $\vert k_R\vert$ that we allow in this approach should be of order the inverse of the patch's radial thickness rather than zero, but at such large $\vert n_\varphi \vert$ the local wavevector $\bk$ will be dominated by $k_\varphi$ anyway, meaning we do not commit a major error by allowing $k_R$ to pass through zero. (These are the same assumptions used when constructing local `shearing sheet' models of galactic disks, in which we take a local patch of a disk and assume it can be treated as infinite and curvature-free --- see \citealt{julian1966non,binney2020angle}.) Again, all we are really saying by making the WKB ansatz \eqref{eqn:WKB} is that the gravitational potential is determined locally within each patch and that the structure of the disk outside the patch can be ignored.}

\red{Let us therefore} insert \eqref{eqn:WKB} into \eqref{eq:potential_expansion_epicyclic}. \red{We soon arrive at}
   \red{\begin{align}
    \delta \phi_{\bn}(\bJ) = u_{n_\varphi}(k_R)\,\me^{ik_R R_\mathrm{g}}
    \sum_{\ell=-\infty}^\infty \sum_{\ell'=-\infty}^\infty 
    (-i)^{n_R-\ell-2\ell'}
    J_\ell\left( n_\varphi \gamma \frac{a_R}{R_\mathrm{g}} \right)
    J_{\ell'}\left( \frac{n_\varphi a_R^2}{4} \frac{\md \kappa}{\md J_\varphi} \right)
    J_{n_R-\ell-2\ell'}(k_Ra_R).
    \label{eqn:WKB_first}
\end{align}} 
\red{Now, the argument of the $J_{\ell'}$ Bessel function is $\sim \vert n_\varphi\vert \epsilon_R^2$. Provided we limit ourselves to $\vert n_\varphi\vert \sim 1$ or $\vert n_\varphi \vert \sim \epsilon_R^{-1}$,  we can again use the approximation \eqref{eqn:Bessel_small_argument} which tells us that only $\ell' = 0,\pm 1$ contribute to this sum up to corrections $\mathcal{O}(\epsilon_R^2)$. Having thus dealt with the $\ell'$ sum we can also carry out the $\ell$ sum using Graf's addition theorem for Bessel functions \citep{abramowitz1968handbook}, which tells us that 
for arbitrary real numbers $u$ and $v$,
\begin{align}
    \sum_{\ell=-\infty}^\infty J_{\ell+\nu}(u)J_{\ell}(v)\me^{i\ell\alpha}=J_\nu(w)\me^{i\nu \chi},
\end{align}
where
\begin{equation}
    w\equiv \mathrm{sgn}({u})\,\sqrt{u^2+v^2-2uv\cos\alpha},
    \,\,\,\,\,\,\,\,\,\,\,\,\,\,\,\,\,\,\,\,\,\,\,\, \chi = \arctan\left(\frac{v}{u-v\cos\alpha} \right).
\end{equation}
Applying this to \eqref{eqn:WKB_first} gives}
\red{\begin{align}
    \delta \phi_{\bn}(\bJ) = u_{n_\varphi}(k_R)\me^{i(k_R R_\mathrm{g}  + n_R\beta)} (-i \, \mathrm{sgn}\,  k_R)^{n_R} \bigg\{J_{n_R}\left(K_R a_R\right)  -\frac{n_\varphi a_R^2}{8} \frac{\md \kappa}{\md J_\varphi}  \left[ \me^{-2i \beta} J_{n_R-2}\left(K_R a_R\right) - \me^{2i \beta} J_{n_R+2}\left(K_R a_R\right) \right]\bigg\},
    \label{eqn:deltaPhi_WKB_general}
\end{align}}
where we defined the modified wavenumber $K_R > 0$ and angle \red{$\beta \in (-\pi/2,\pi/2)$}:
\begin{equation}
    K_R \equiv \sqrt{k_R^2 + \frac{n_\varphi^2\gamma^2}{R_\mathrm{g}^2}} = \vert k_R \vert \sqrt{1 + \gamma^2 \bigg\vert \frac{k_\varphi}{k_R} \bigg\vert^2},\,\,\,\,\,\,\,\,\,\,\,\,\,\,\,\,\,\,\,\,\,\,\,\,\,\,\,\,\,
    \label{eqn:modified_k}
   \red{\beta \equiv \arctan\left( \frac{ n_\varphi  \gamma}{ k_R  R_\mathrm{g}} \right) = \arctan\left( \frac{\gamma k_\varphi}{k_R}  \right),}
\end{equation}
both of which depend on $k_R$, $n_\varphi$ and $J_\varphi$, but not on $J_R$. 

\red{The `Dehnen' terms in the curly bracket in \eqref{eqn:deltaPhi_WKB_general} are of order $\sim n_\varphi \epsilon_R a_R^2 \vert \md\kappa/\md J_\varphi \vert$ (where we have replaced the factor $1/8$ by $\sim \epsilon_R$). For smooth rotation curves we normally have $\vert \md \kappa/\md J_\varphi\vert \sim R_\mathrm{g}^{-2}$, meaning this term is $\mathcal{O}(\vert n_\varphi \vert \epsilon_R^3 )$, which is ignorable for any $\vert n_\varphi\vert \lesssim \epsilon_R^{-1}$. It is also ignorable for even larger $\vert n_\varphi \vert$ if for some reason $\md \kappa /\md J_\varphi$ is unusually small or zero, as is assumed when constructing the shearing sheet \citep{julian1966non,binney2020shearing}. In either of these cases we get\footnote{\red{However there are scenarios in which $\md \kappa / \md J_\varphi$ can become large, such as in the presence of disk breaks (see e.g., Figure 3 of \citealt{fiteni2024role}), in which case one should reconsider the approximations made in moving from \eqref{eqn:deltaPhi_WKB_general} to \eqref{eqn:deltaPhi_WKB}.}} 
\begin{align}
    \delta \phi_{\bn}(\bJ) = u_{n_\varphi}(k_R)\me^{i(k_R R_\mathrm{g}  + n_R\beta)} (-i \, \mathrm{sgn}\,  k_R)^{n_R} J_{n_R}\left(K_R a_R\right).
    \label{eqn:deltaPhi_WKB}
\end{align}}
\red{In fact, if $\vert n_\varphi \vert \sim 1$ then in order to be in the short wavelength regime we must have $\vert k_R \vert \sim R_\mathrm{g}/a_R^2$. It follows that short wavelength fluctuations with $\vert n_\varphi \vert \sim 1$  are necessarily \textit{tightly wound}, by which we mean
\begin{equation}
    \bigg \vert \frac{k_\varphi}{k_R} \bigg \vert \sim \epsilon_R^2.
\end{equation}}
Then in \eqref{eqn:deltaPhi_WKB} \red{we may replace $K_R \simeq \vert k_R\vert$ and $\beta \simeq 0$ up to corrections $\mathcal{O}(\epsilon_R^2)$, thereby arriving at equation \eqref{eqn:very_short_perturbations}.}

\red{Strictly, we should not use \eqref{eqn:deltaPhi_WKB_general} at all if $\vert n_\varphi \vert \sim \epsilon_R^{-2}$.
The reason is that  the argument of the $J_{\ell'}$ Bessel function in \eqref{eqn:WKB_first} is then of order unity, so we cannot rely on the expansion \eqref{eqn:Bessel_small_argument} to truncate the sum at $\vert \ell' \vert \leq 1$; if we did truncate there, the error in the result \eqref{eqn:deltaPhi_WKB} would be formally $\mathcal{O}(\epsilon_R)$ rather than our desired $\mathcal{O}(\epsilon_R^2)$.
This issue is exacerbated further if we go to even higher azimuthal wavenumbers, e.g., $\vert n_\varphi \vert \gtrsim \epsilon_R^{-3}$; in general, by making $\vert n_\varphi\vert$ arbitrarily large we find ourselves having to retain arbitrarily many $\ell'$ values in the sum in \eqref{eqn:WKB_first}.
Physically, the problem is that $\delta \phi_{\bn}(\bJ)$ corresponds to a fluctuation field integrated over a star's epicyclic orbit (see equation \eqref{eq:xxx}), 
and when this field has arbitrarily small-scale features, minor corrections to the epicyclic approximation (including, but not limited to, the Dehnen drift) become arbitrarily important. The lesson we take from this is that at such short wavelengths, $\delta \phi_{\bn}$ is not a particularly useful object at all --- in practice, it turns out to be better to treat the fluctuation field as a local set of plane waves $\me^{i\bk\cdot\br}$ and perform calculations first in position-velocity space rather than angle-action space.
This is why we said below equation \eqref{eqn:def_short_wavelength_regime} that we would not consider fluctuations with $\vert k_R \vert \gtrsim R_\mathrm{g}^2 a_R^{-3}$ or  $\vert n_\varphi \vert \sim \epsilon_R^{-3}$. Given the above discussion we should also refuse to consider $\vert n_\varphi \vert \sim \epsilon_R^{-2}$, but since we are only committing a small crime in this case we will proceed as if $\vert n_\varphi \vert \sim \epsilon_R^{-2}$ fluctuations could also be described by \eqref{eqn:deltaPhi_WKB} to the accuracy we desire.}

\bibliography{sample}{}
\bibliographystyle{aasjournal}

%% This command is needed to show the entire author+affiliation list when
%% the collaboration and author truncation commands are used.  It has to
%% go at the end of the manuscript.
%\allauthors

%% Include this line if you are using the \added, \replaced, \deleted
%% commands to see a summary list of all changes at the end of the article.
%\listofchanges

\end{document}